\documentclass[a4paper]{article}
\usepackage{listings}
\usepackage{xcolor}
\usepackage{multirow}
\usepackage{jheppub}  
\usepackage{dcolumn}
\usepackage[english]{babel}
\usepackage[utf8x]{inputenc}
\usepackage[T1]{fontenc}
\usepackage{palatino}
\usepackage{amsmath}
\usepackage{afterpage}
\usepackage{graphicx, subcaption}
\usepackage[colorinlistoftodos]{todonotes}
\usepackage[colorlinks=true]{hyperref}
\usepackage{makeidx}
\newcommand{\be}{\begin{equation}}  
\newcommand{\ee}{\end{equation}}  
\newcommand{\bea}{\begin{eqnarray}}  
\newcommand{\eea}{\end{eqnarray}}  
\makeindex
\begin{document}

\vspace*{1.2cm}

\thispagestyle{empty}
\begin{center}

{\LARGE \bf Review of results on forward physics and diffraction by CMS}

\par\vspace*{7mm}\par

{

\bigskip

\large \bf Cristian Baldenegro \\ (on behalf of the CMS Collaboration)}

\bigskip

{\large \bf  E-Mail: c.baldenegro@cern.ch}

\bigskip

{ University of Kansas}

\bigskip

{\it Presented at the Workshop of QCD and Forward Physics at the EIC, the LHC, and Cosmic Ray Physics in Guanajuato, Mexico, November 18-21 2019}

\vspace*{15mm}

{  \bf  Abstract }

%Recent results

\end{center}
\vspace*{1mm}

\begin{abstract}

Results by the CMS Collaboration on forward physics, diffraction, and physics in the small-$x$ limit of quantum chromodynamics (QCD), are presented. In particular, results on azimuthal angle decorrelations between two jets in events where two outermost jets are separated by a large rapidity interval are discussed. In addition, results based on the production of two jets separated by a large rapidity gap (interval void of radiation) are presented. These dijet production processes are expected to be sensitive to Balitsky-Fadin-Kuraev-Lipatov (BFKL) evolution effects. We highlight results on inclusive forward jet production and on exclusive vector meson production in proton-lead collisions, which access gluon densities in the small-$x$ and $Q^2$ where saturation effects may play a role. A summary of results on underlying event activity studies based on inclusive $Z$ boson production, charged particle spectra in minimum bias events, and energy density in forward pseudorapidities is presented. These studies provide valuable inputs for Monte Carlo event generator tuning, and test predictions based on perturbative and non-perturbative QCD techniques.

\end{abstract}
  
\section{Introduction}

Present particle physics searches rely on collisions of protons at very high energies. We are mostly interested in the interactions of quarks and gluons of the protons. Thus, a fundamental ingredient in our understanding of particle physics in modern colliders relies in our description of the strong interaction, based on quantum chromodynamics (QCD), the quantum theory of strong interactions. At very short distances (much smaller than the size of the proton), we can rely on perturbation theory techniques, where production rates are presented in a power series expansion in the strong coupling constant, $\alpha_s \ll 1$. However, at large distances (larger than the proton size), this is no longer possible, and a phenomenological approach has to be adopted instead. For these reasons, a tremendous amount of work has gone in further understanding the regimes of short and long distance physics in QCD.

Of particular interest is the so-called small-$x$ limit of QCD, where $x$ represents the parton momentum fraction relative to the proton. Indeed, the large density of gluons at small-$x$ can be understood in terms of the myriad of parton splittings occurring at small-$x$ within the proton. The parton density evolution in the small-$x$ limit is described by the Balitsky-Fadin-Kuraev-Lipatov (BFKL) evolution equation. The latter predicts a power-law growth of parton densities at low $x$. Smoking-gun experimental evidence for BFKL dynamics has yet to be found. On the other hand, the parton cascade described by BFKL evolution leads to a violatfion of unitarity at very small values of $x$. Non-linear evolution of parton distribution functions (PDFs), which incorporate gluon-gluon recombination mechanism, may play a role in this regime. Said mechanism is believed to slow down the rapid growth of the proton structure function at very small values of the parton momentum fraction $x$. So far, experimental evidence in support of parton saturation effects has not yet been found.

In addition, we want to refine our understanding of the underlying dynamics in low-momentum exchange processes in hadronic collisions. The description of these phenomena rely on phenomenological models whose parameters are ``tuned'' to data. Dedicated measurements sensitive to soft physics effects provide valuable inputs for Monte Carlo (MC) event generators. The latter are widely used for precision studies of standard model processes, searches for new physics, and in cosmic-ray physics.

In this report, we present a selected number of results by the CMS Collaboration~\cite{CMS} related to forward physics, small-$x$, and diffraction. These were presented at the QCD and Forward Physics meeting held in Guanajuato, Mexico from November 18 through November 21st 2019.

\section{Mueller-Navelet jets at 7 TeV}

\begin{figure}[]
\begin{center}
\includegraphics[width=.49\textwidth]{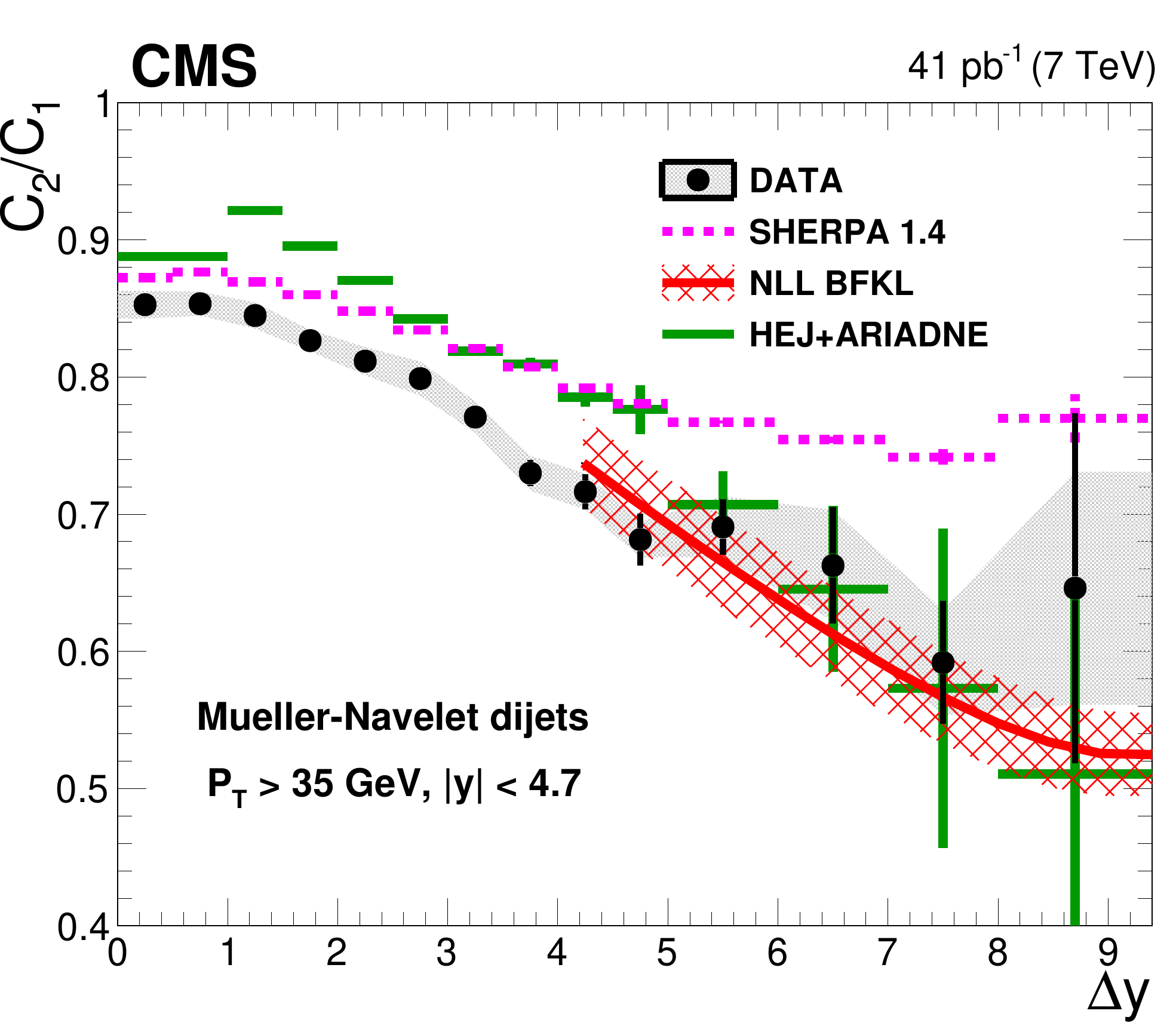}
\includegraphics[width=.49\textwidth]{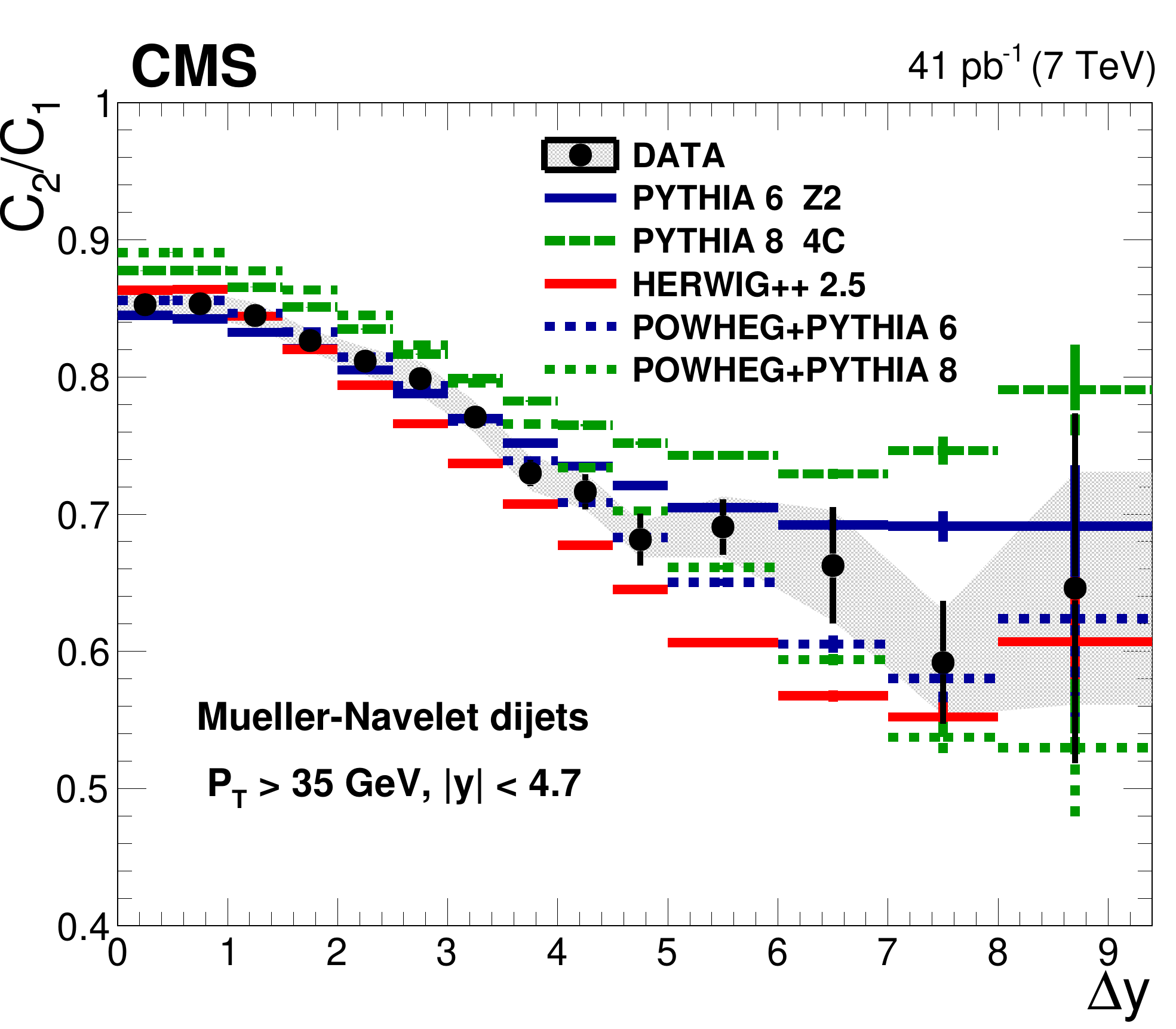}
\includegraphics[width=.49\textwidth]{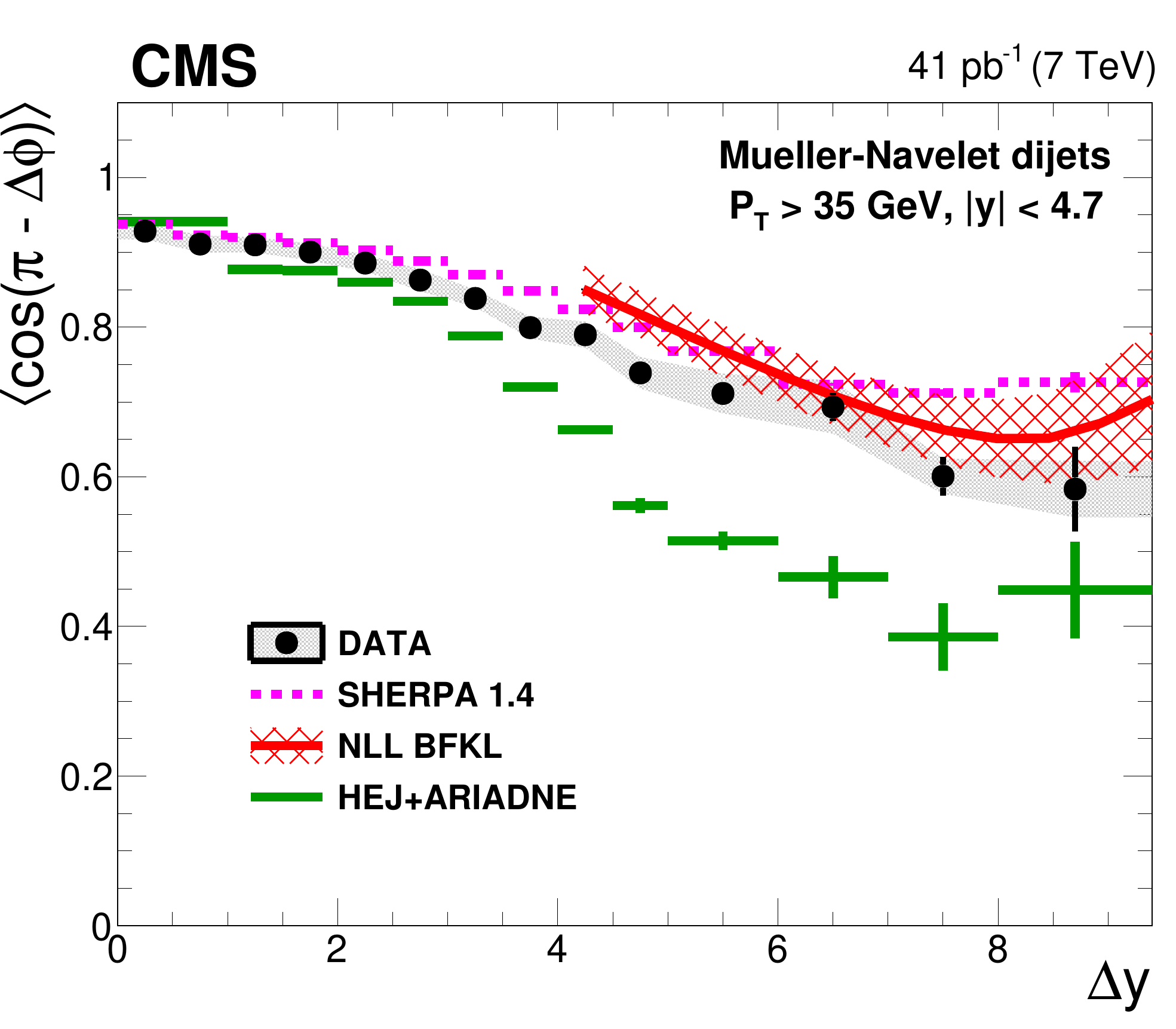}
\includegraphics[width=.49\textwidth]{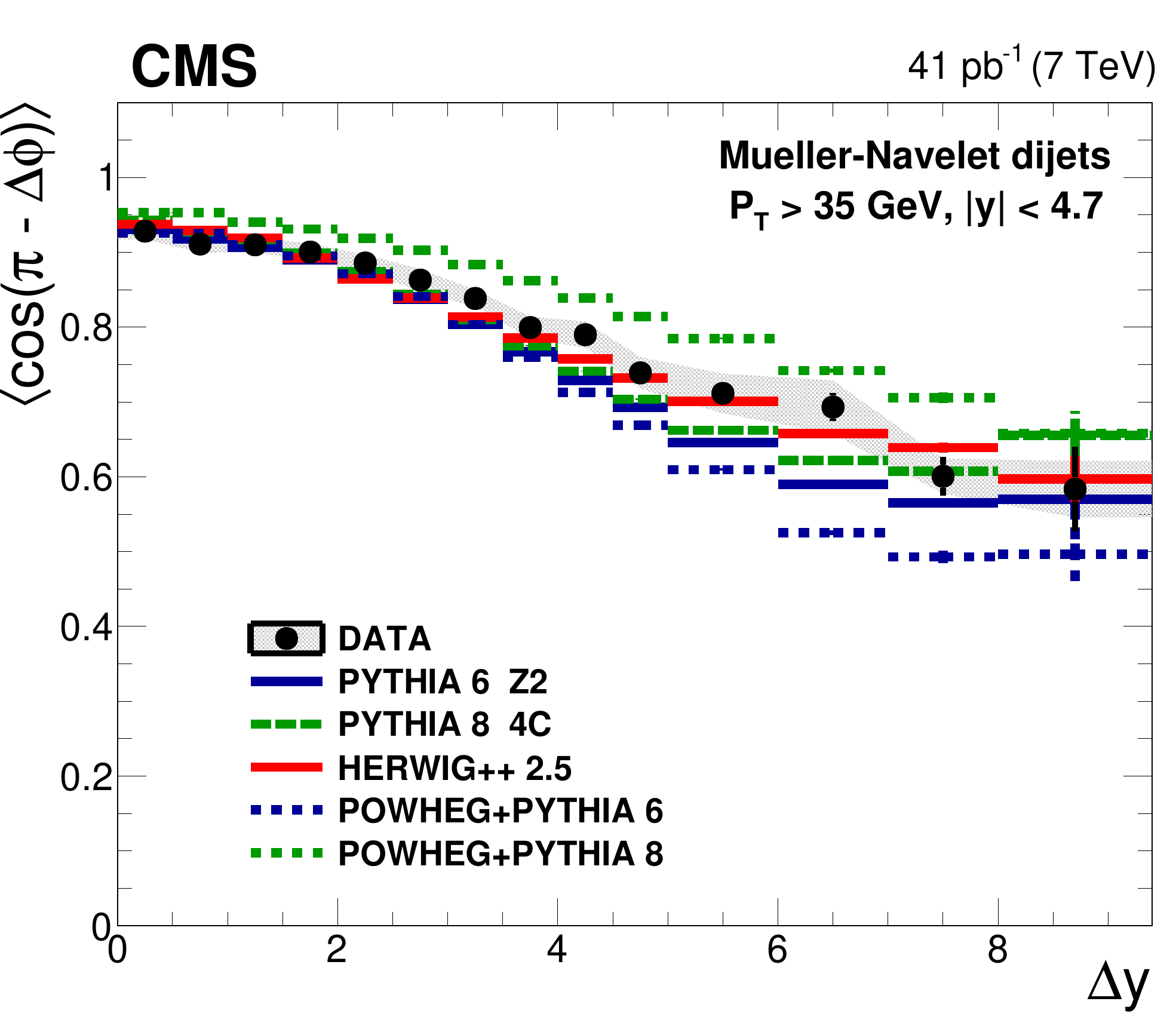}
\end{center}
\caption{\label{fig:MN_jets} (Top) Ratio of the coefficients $C_2$ and $C_1$ as a function of the rapidity difference between the outermost jets, $\Delta y$. The coefficients $C_n$ are the average cosine of $n(\pi-\Delta\phi)$. CMS results are represented by the data points. (Bottom) Average cosine of the azimuthal angle separation between the outermost jets as a function of the rapidity difference $\Delta y$ between the jets. Predictions based on BFKL-NLL calculations are represented by the red band. Predictions based on DGLAP and BFKL-LL MC generators are represented by the other curves, as described in text. Figures are extracted from Ref.~\cite{mn_CMS}.}
\end{figure}

BFKL dynamics can be probed in the production of two jets separated by a large interval in rapidity. This configuration is known as Mueller-Navelet jets. The parton cascade described in the BFKL approach would induce angular decorrelations between the final-state jets. These decorrelations are expected to be stronger than those expected from the parton cascade described by Dokshitzer-Gribov-Lipatov-Altarelli-Parisi (DGLAP) evolution, where parton emissions are strongly ordered in transverse momentum.

The CMS experiment reported a study on Mueller-Navelet jets based 7 TeV pp data~\cite{mn_CMS}. The number of overlapping pp collisions within a single bunch crossing (``pileup'' events) was low for the data used in this analysis. The selection criteria considers two jets with $p_T > 35$ GeV reconstructed with the anti-$k_t$ algorithm~\cite{Cacciari_2008} with distance parameter $R = 0.5$ within the full acceptance of CMS, $|y|<4.7$. The key observable here is the distribution of the azimuthal angle separation between the jets, $\Delta\phi \equiv |\phi_\text{jet1}-\phi_\text{jet2}|$, normalized to unity, in bins of $\Delta y\equiv |y_\text{jet1} - y_\text{jet2}|$. Based on these distributions, one can extract the average cosines of $(\pi - \Delta \phi)$, $2(\pi-\Delta\phi)$, $3(\pi-\Delta\phi)$, and ratios of these average cosines.

Predictions based on various MC event generators based on DGLAP evolution, on BFKL evolution with resummation of logarithms of energy at leading-logarithm (LL), and analytical prediction based on BFKL evolution at next-to-leading-logarithm (NLL) accuracy are presented for these observables. BFKL at NLL calculations describe data at large $\Delta y$ within uncertainties. HEJ+ARIADNE~\cite{ariadne}, based on LL BFKL amplitudes, underestimates data at large $\Delta y$. PYTHIA8~\cite{pythia8}, HERWIG++~\cite{herwigpp}, SHERPA~\cite{sherpa}, based on leading order (LO) DGLAP calculation, are able to describe data over wide range in $\Delta y$ within uncertainties. POWHEG next-to-leading order (NLO) predictions supplemented with PYTHIA6~\cite{pythia6} or PYTHIA8 for parton shower and hadronization effects underestimates or overestimates data at large $\Delta y$, respectively. With the present observables and experimental uncertainties, it was found that both BFKL or DGLAP based approaches are able to describe the data. Extensions of this study that account for the interjet activity may be able to isolate BFKL dynamics better, in addition to these observables based on the $\Delta\phi$ distributions.

\section{Jet-gap-jet at 7 TeV}

Another process that is potentially highly sensitive to BFKL dynamics is the production of two jets separated by a large rapidity gap, where the rapidity gap is an interval void of particle activity. This is known as jet-gap-jet or Mueller-Tang jets. Here, DGLAP dynamics are heavily suppressed due to the rapidity gap requirement by way of a Sudakov form factor. In jet-gap-jet events, contributions from color-singlet exchange are largely favored. In presence of a hard energy scale (given by the $p_T$ of the jets), the perturbative pomeron exchange is the preferred mechanism in strong interactions to generate the rapidity gap. The perturbative pomeron exchange consists of two-gluon ladder exchange in the BFKL framework.

CMS has presented results on jet-gap-jet events in low pileup $7$ TeV pp collisions~\cite{jgj_7TeV_CMS}. In said study, anti-$k_t$ particle-flow jets of distance parameter of $R = 0.5$ are considered. The leading two jets are required to have $p_T^\text{jet} > 40$ GeV and $1.5<\eta_\text{jet}|<4.7$ each, with opposite signed pseudorapidities $\eta_\text{jet1}\times\eta_\text{jet2} < 0$. The rapidity gap is defined by means of the charged particle multiplicity in $|\eta|<1$, where each charged particle has $p_T > 200$ MeV. Data-based methods are used to estimate the contribution of color exchange dijet events with downwards fluctuations in the particle multiplicity, such that they yield a rapidity gap. The latter are subtracted from the data in order to extract the color-singlet exchange contribution.

\begin{figure}[ht!]
\begin{center}
\includegraphics[width=.49\textwidth]{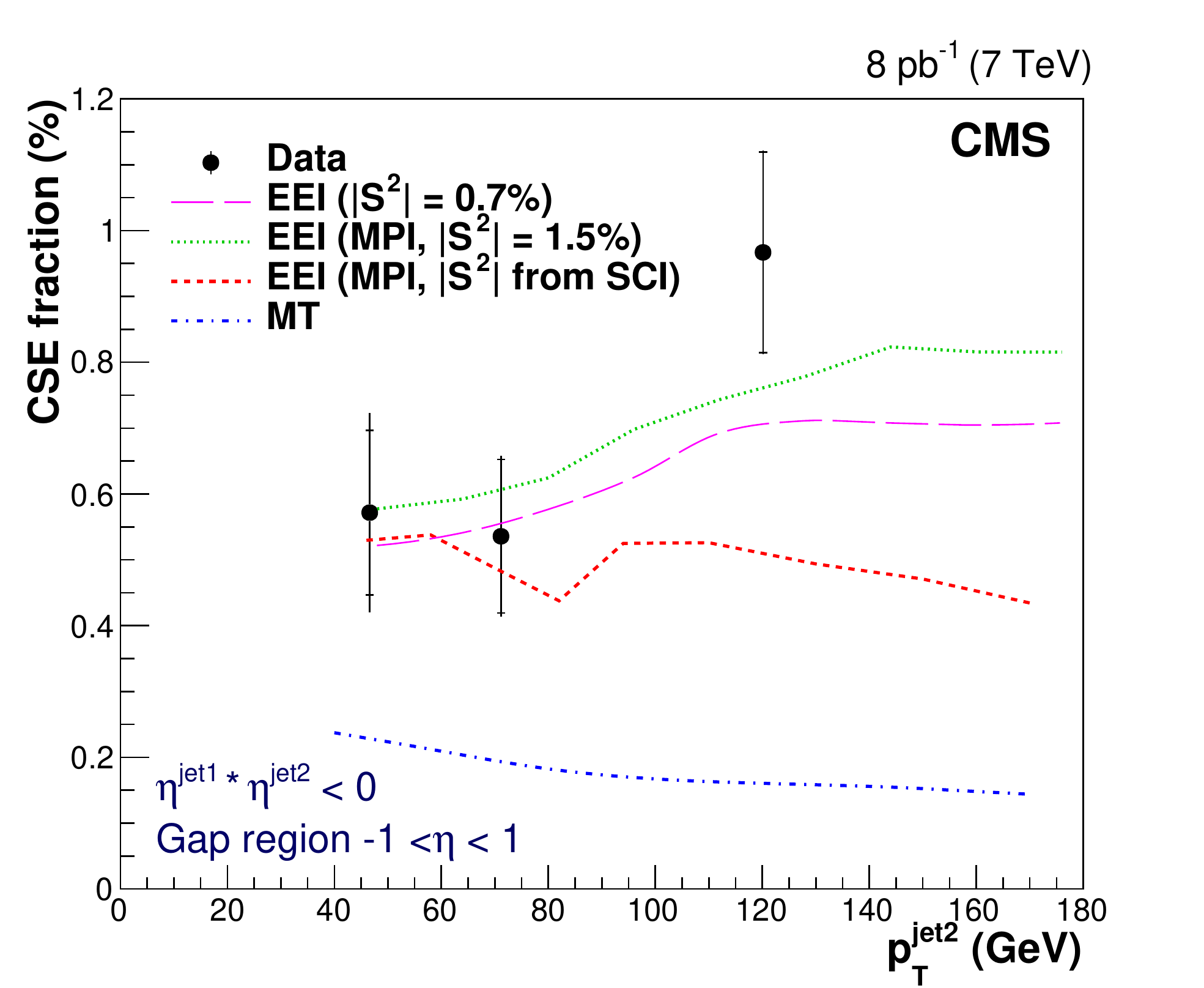}
\includegraphics[width=.49\textwidth]{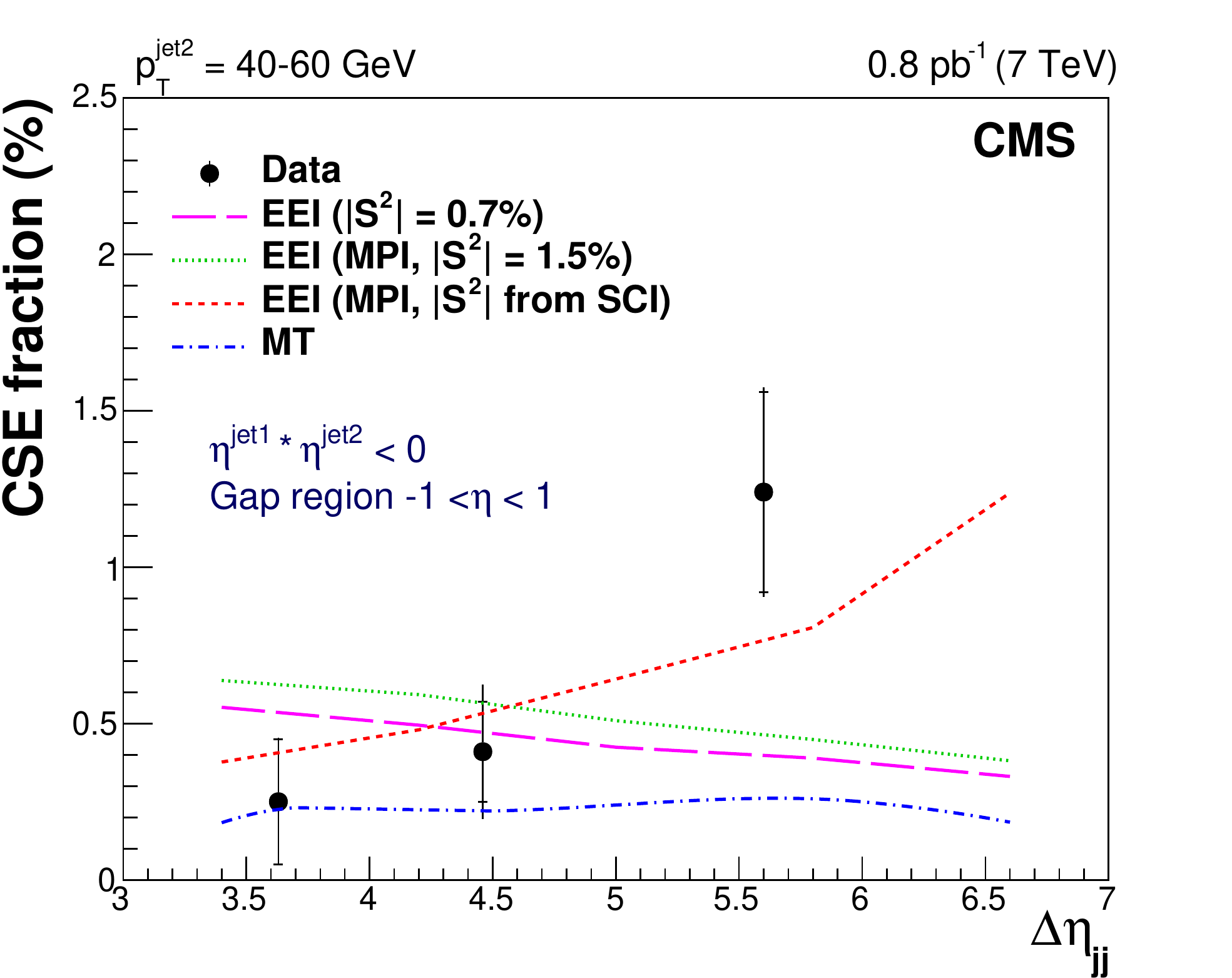}
\end{center}
\caption{\label{fig:jet-gap-jet} Fraction of color-singlet exchange dijet events, $f_\text{CSE}$, as a function of the subleading jet transverse momentum, $p_{T}^\text{jet2}$ (Left) and pseudorapidity difference between the jets, $\Delta\eta_{jj}$ (Right). Predictions based on BFKL calculations based on the IEE model. Figures are extracted from Ref.~\cite{jgj_7TeV_CMS}.}
\end{figure}

The ratio of events identified as originating from color-singlet exchange divided by the total number of dijet events is measured as a function of the subleading jet $p_T^\text{jet2}$ and the pseudorapidity difference between the jets $\Delta\eta_\text{jj}$, as shown in Fig.~\ref{fig:jet-gap-jet}. Predictions based on BFKL NLL calculations supplemented with survival probability effects based on multiple partonic interactions (MPI) or the soft color interaction model (SCI) by Ingelman-Ekstedt-Enberg (IEE)~\cite{iee} are shown in Fig.~\ref{fig:jet-gap-jet}. The predictions based on BFKL NLL calculations are not able to simultaneously describe the results in $p_T^\text{jet2}$ and $\Delta\eta_\text{jj}$. A larger dataset at a higher $\sqrt{s}$ may provide additional insight into the nature of the mechanism behind rapidity gap generation in these events. Since these results were presented in this meeting, preliminary results by the CMS and TOTEM Collaborations on jet-gap-jet production at 13 TeV have been publicly presented, as shown in Ref.~\cite{jetgapjet_13TeV}. The corresponding paper is in preparation.

\section{Very forward jet production cross section in pPb collisions}

The CMS experiment is equipped with the CASTOR electromagnetic and hadronic calorimeter~\cite{CMS}, which extends the measurable jets pseudorapidity acceptance up to $-6.6 < \eta  < -5.2$ with approx. $p_T\geq 4$ GeV. A study of jets in CASTOR in p+Pb collisions possesses unique sensitivity to non-linear evolution effects due to the very forward acceptance for jet reconstruction ($x\sim 10^{-6}$) and the enhancement of the parton densities in the ion (scales with the number of nucleons as $A ^{1/6}$).

\begin{figure}[ht!]
\begin{center}
\includegraphics[width=.49\textwidth]{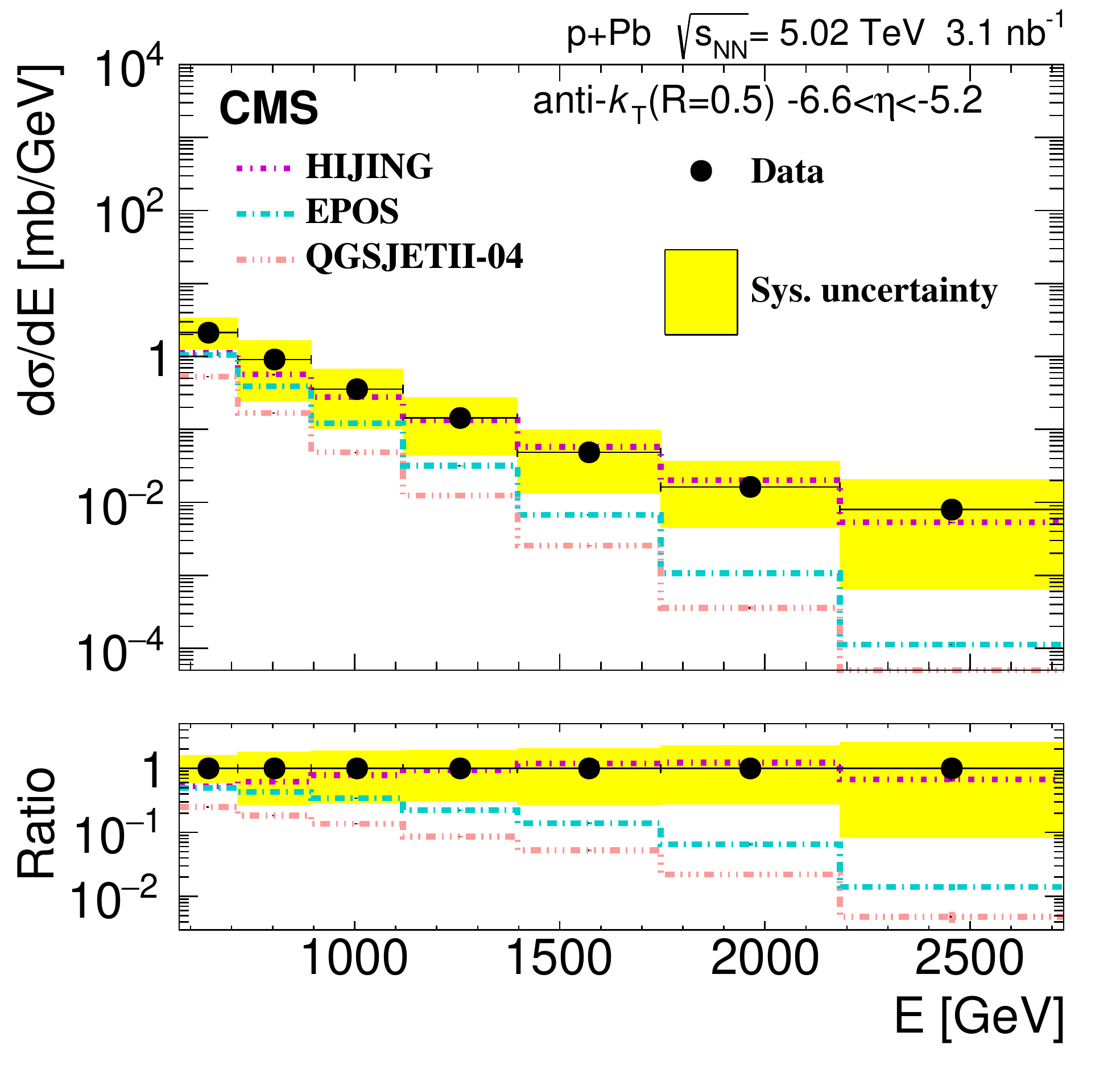}
\includegraphics[width=.49\textwidth]{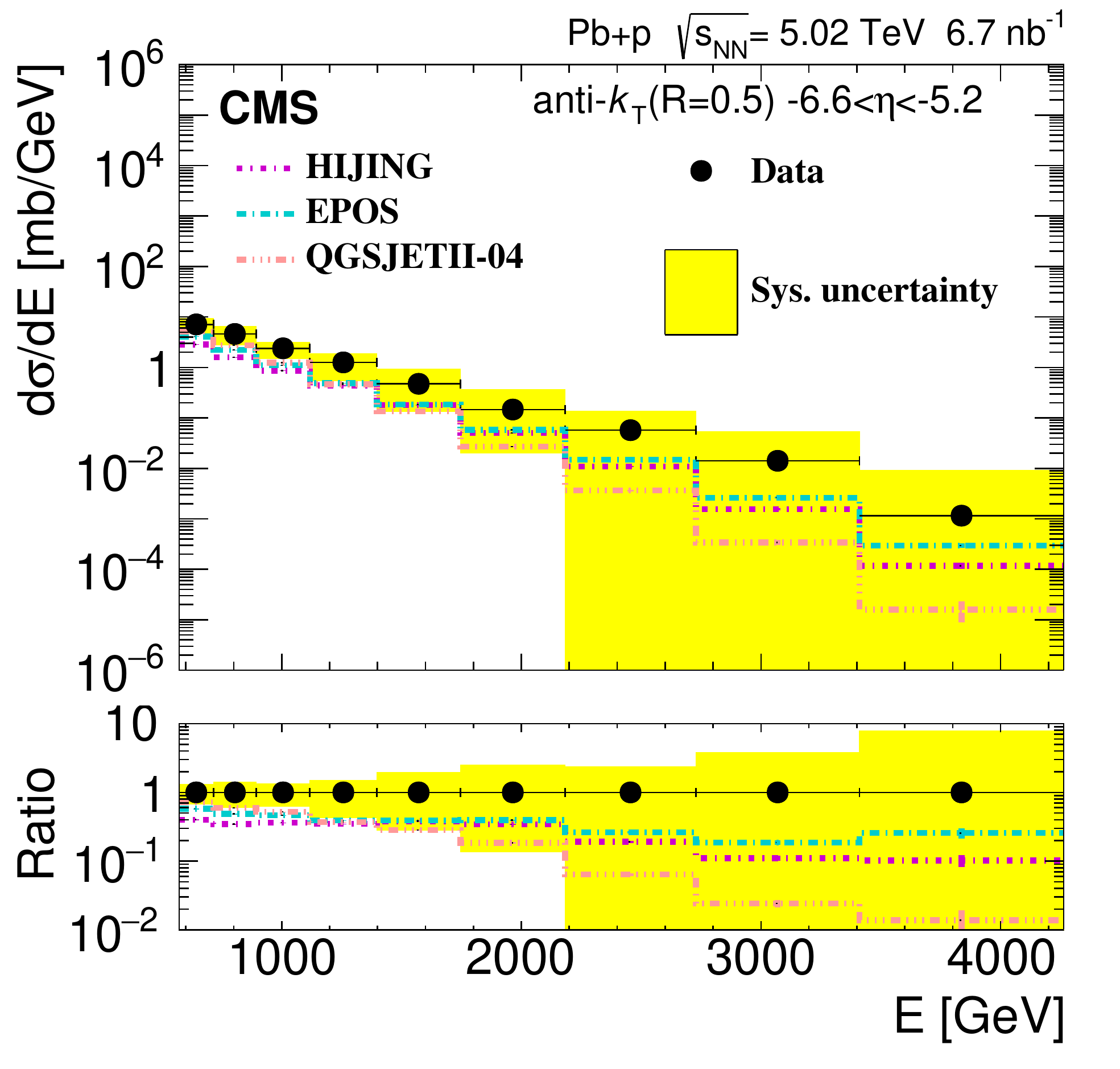}
\includegraphics[width=.49\textwidth]{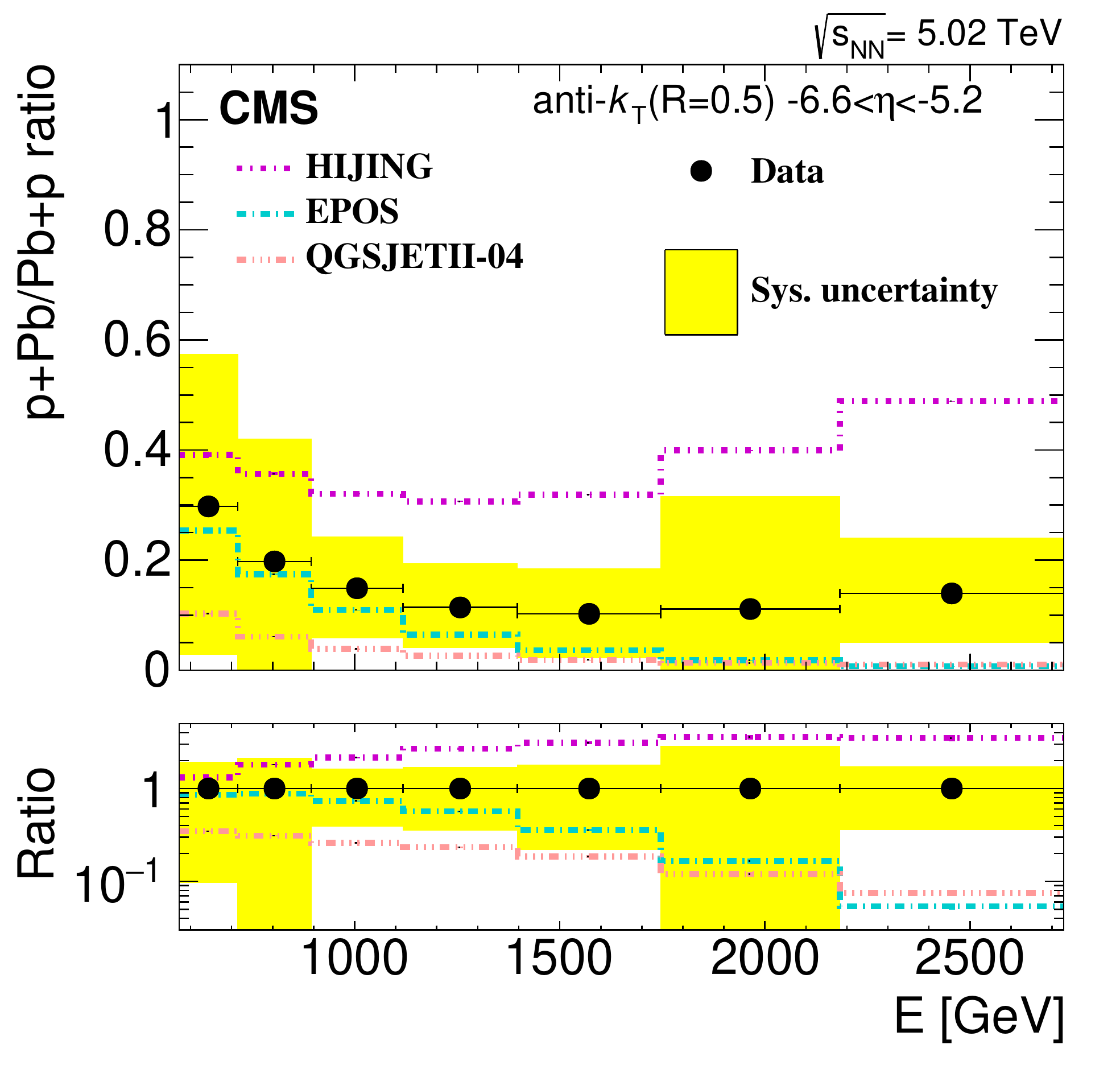}
\end{center}
\caption{\label{fig:fwd_jets} Differential cross section as a function of the jet energy in CASTOR for the p+Pb configuration (left), the Pb+p configuration (right), and the ratio of the differential cross section in p+Pb to Pb+p as a function of the jet energy(bottom). Predictions based on HIJING, EPOS-LHC, and QGSJETII are shown on top of the data. Figures are extracted from Ref.~\cite{forward_jets_CMS}.}
\end{figure}

The measurement of forward inclusive jet cross section is done for the p+Pb  (proton towards CASTOR) and the Pb+p (Pb-ion towards CASTOR) configurations. The p+Pb configuration is ideal to probe small-$x$ physics~\cite{forward_jets_CMS}. The sample was collected in pPb collisions at a center-of-mass energy per nucleon pair of $\sqrt{s}_\text{NN} = 5.02$ TeV in the laboratory frame (proton beam energy of 4 TeV and 1.6 TeV per nucleon for the Pb beam) using a minimum bias trigger. To suppress the contribution from diffractive and photon-induced processes, a requirement of minimum one calorimeter tower with energy above 4 GeV in $3<|\eta|<5$ on both sides is applied. The main result of the study is the differential cross section as a function of the jet energy deposited in CASTOR. The cross sections  are unfolded to particle level jets, and are shown in Fig.~\ref{fig:fwd_jets}. The leading systematic uncertainty comes from the CASTOR jet energy scale, followed by uncertainties associated to the model dependence from the unfolding procedure and the alignment and calibration corrections. HIJING~\cite{hijing}, EPOS~\cite{epos} and QGSJetII~\cite{qgsjet2} event generators are compared to data; they each have a different treatment for non-linear evolution effects of the parton PDFs, as described in Ref.~\cite{forward_jets_CMS}. HIJING predictions are in agreement with data in the p+Pb configuration, while EPOS and QGSJetsII progressively underestimate the cross section with increasing energy. In the Pb+p configuration, HIJING and EPOS give a reasonable description of the shape of the distribution found in data. To further enhance the model discrimination power in the analysis, the ratio of the p+Pb and Pb+p spectra is measured as a function of the jet energy. The ratio approximately cancels the jet energy scale uncertainties, leaving the model dependence from the unfolding procedure as the leading systematic uncertainty. This observable has the caveat that the ratio is performed on proton-lead collision configurations boosted with respect to each other; therefore, the same values of the parton momentum fraction $x$ is not probed in both configurations simultaneously. HIJING  describes the shape well, but the normalization disagrees by a factor of 2. EPOS and QGSJetII are off in shape and show a large discrepancy at increasingly large energies. Other predictions that treat the low-$x$ gluon densities differently than the aforementioned models are described in Ref.~\cite{forward_jets_CMS}.

\section{Exclusive vector meson production in ultraperipheral pPb collisions}

At the CERN LHC, one can use ultraperipheral proton-lead collisions to study exclusive vector meson production. Indeed, the electromagnetic field generated by the relativistic lead-ion can be treated as a source of quasi-real photons. The photon luminosity receives an enhancement proportional to the square of the number of protons in the nucleus, $Z^2$. The quasi-real photons emitted by the lead nucleus can then probe the proton, and if the interaction is hard enough, it can probe the parton densities of the proton. Due to the asymmetry of the type of beams, one can unambiguously identify the direction of the photon emission. This renders a situation similar to that of electron-proton collisions at HERA.

Diffractive photoproduction of quarkonia offers a clean probe of gluon densities of the proton at small values of $x = 10^{-4}$--$10^{-2}$ and small values of $Q^2 \approx m_V^2$. In these interactions, the quasi-real photon emitted by the lead nucleus fluctuates into a vector meson, which probes the gluon PDFs of the proton via two-gluon exchange (pomeron exchange). At LO in perturbative QCD (pQCD), the production rate is proportional to the square of the gluon PDF. The latter suggests that exclusive production of vector mesons is highly sensitive to physics effects that may take place at low-$x$, such as parton saturation effects.

The CMS experiment has presented results on exclusive $\Upsilon(\text{nS})$ ($n =1, 2, 3$) bottomonium meson production in proton-lead collisions at $\sqrt{s}_{NN} = 5.02$ TeV~\cite{upsilon_CMS}. The rest mass of the $\Upsilon$ meson is large enough (of the order of 10 GeV) to have stable pQCD calculations, but is small enough such that saturation effects may still play an important role in the small-$x$ regime of gluon densities. In the study, $\Upsilon$ meson decays in muon pairs in the rapidity range $|y|<2$ are considered.

\begin{figure}[ht!]
\begin{center}
\includegraphics[width=.49\textwidth]{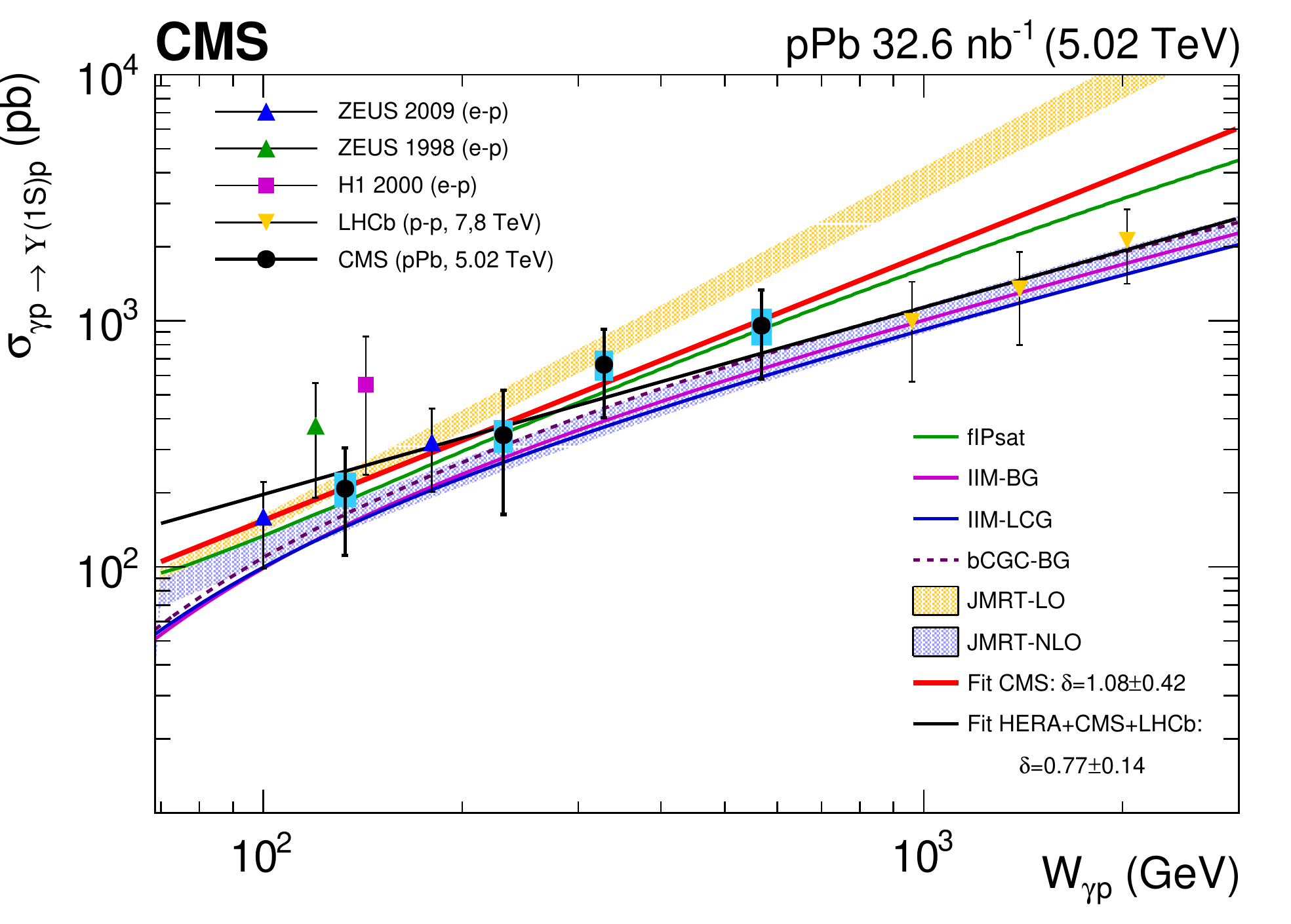}
\includegraphics[width=.49\textwidth]{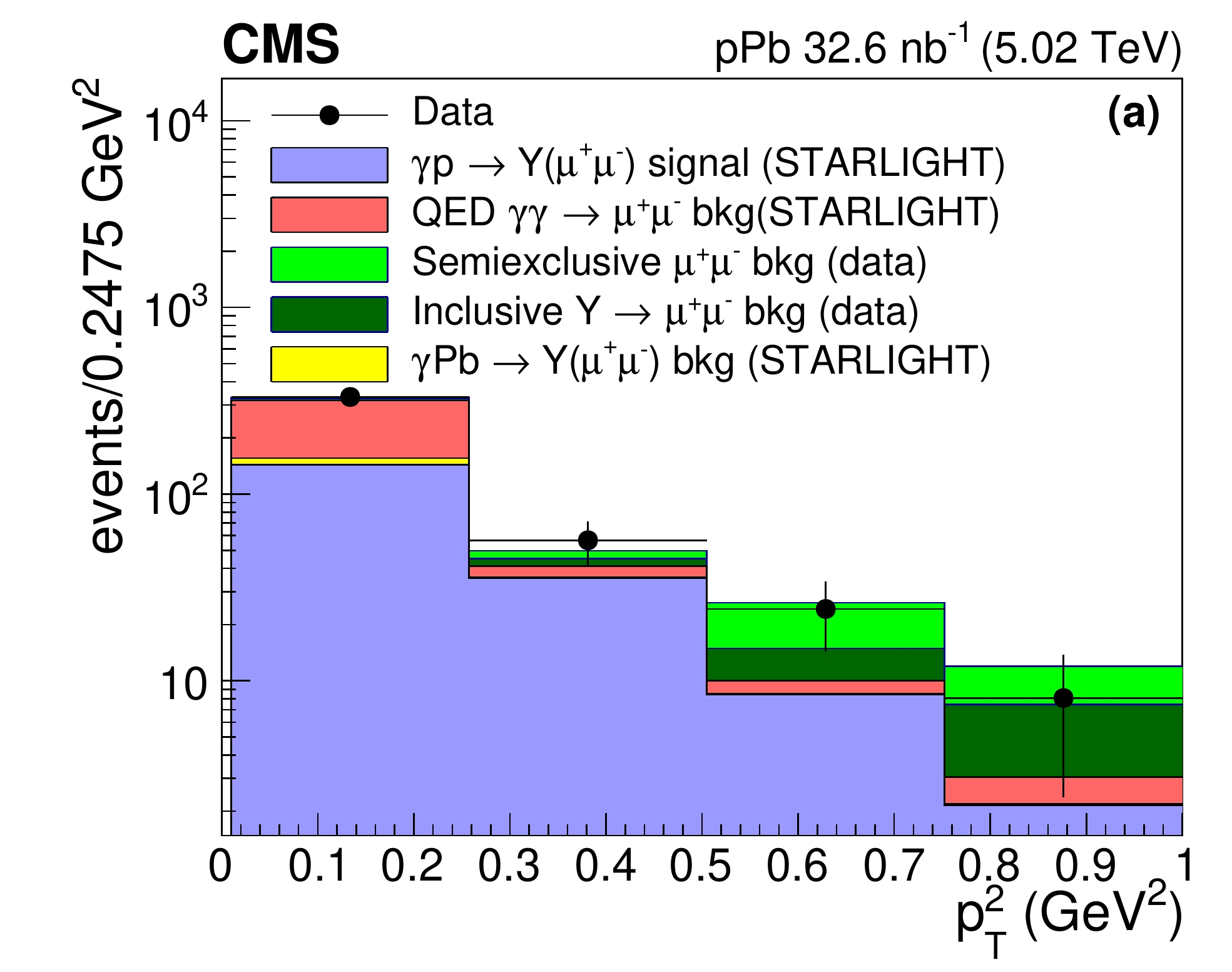}
\end{center}
\caption{\label{fig:upsilon} (Left) Photoproduction cross section as a function of the photon-proton center-of-mass energy, $W_{\gamma p}$. Previous measurements by the H1, ZEUS and LHCb Collaboration are shown as well. (Right) Number of events as a function of the transverse momentum of exclusive $\Upsilon$ meson event candidates, together with background and signal contributions. Figures are extracted from Ref.~\cite{upsilon_CMS}.}
\end{figure}

Upon subtraction of photoproduction of lepton pairs, and non-diffractive $\Upsilon$ production, one can reconstruct the $p_T$ of the extracted exclusive $\Upsilon$ candidates. The latter is a good approximation to the four-momentum transfer square at the proton vertex, and can be seen in Fig.~\ref{fig:upsilon}. The resulting distribution is fit with an exponential function $\exp (-b p_T^2)$. The slope is found to be $b = 6.0 \pm 2.1$ (stat) $\pm 0.3$ (syst) GeV$^{-2}$, in agreement with the value measured by ZEUS at lower photon-proton masses. The slope of the distribution gives information on the gluon PDFs in impact parameter space. The photon-proton center-of-mass energy $W_{\gamma p}$ can be deduced from the rapidity of the $\Upsilon$ in the laboratory frame via $W_{\gamma p}^2 = E_p m_{\Upsilon} \exp (\pm y)$, where $E_p = 4$ TeV. The photoproduction cross section $\sigma(\gamma p\rightarrow \Upsilon p)$ is extracted from $\frac{d \sigma}{d y} (\text{pPb}\rightarrow \text{p}\Upsilon (1S) \text{Pb})$ in bins of $\langle y \rangle$, based on the photon spectrum embedded in STARLIGHT~\cite{starlight}. The CMS pPb results cover a region unexplored by H1, ZEUS, and LHCb. Predictions that account for different $\Upsilon$ wave function parametrizations or small-$x$ gluon PDF models are shown in Fig.~\ref{fig:upsilon}.

CMS has also presented results in exclusive $\rho(770)^0$ photoproduction from protons in pPb collisions at $\sqrt{s}_{NN} = 5.02$ TeV~\cite{rho_CMS}. The smaller mass of the $\rho$ meson corresponds to a larger effective color-dipole size to probe the proton. This renders better sensitivity of possible parton saturation effects. The analysis strategy is very similar to that of exclusive $\Upsilon$ meson production described above. In the analysis, decays of the $\rho(770)^0 \rightarrow \pi^+\pi^-$ are considered. The $p_T$ of the leading and subleading pions is of at least 0.4 and 0.2 GeV, respectively, within $|\eta|<2$. Here, the dominant backgrounds correspond to resonant and non-resonant $\pi^+\pi^-$ (simulated with the STARLIGHT event generator~\cite{starlight}) production, photoproduction of $\rho^0(770)$ with proton dissociation (based on normalization at large $p_T^{\pi^+\pi^-}$), and $\rho(1700)$ production.

\begin{figure}[ht!]
\begin{center}
\includegraphics[width=.54\textwidth]{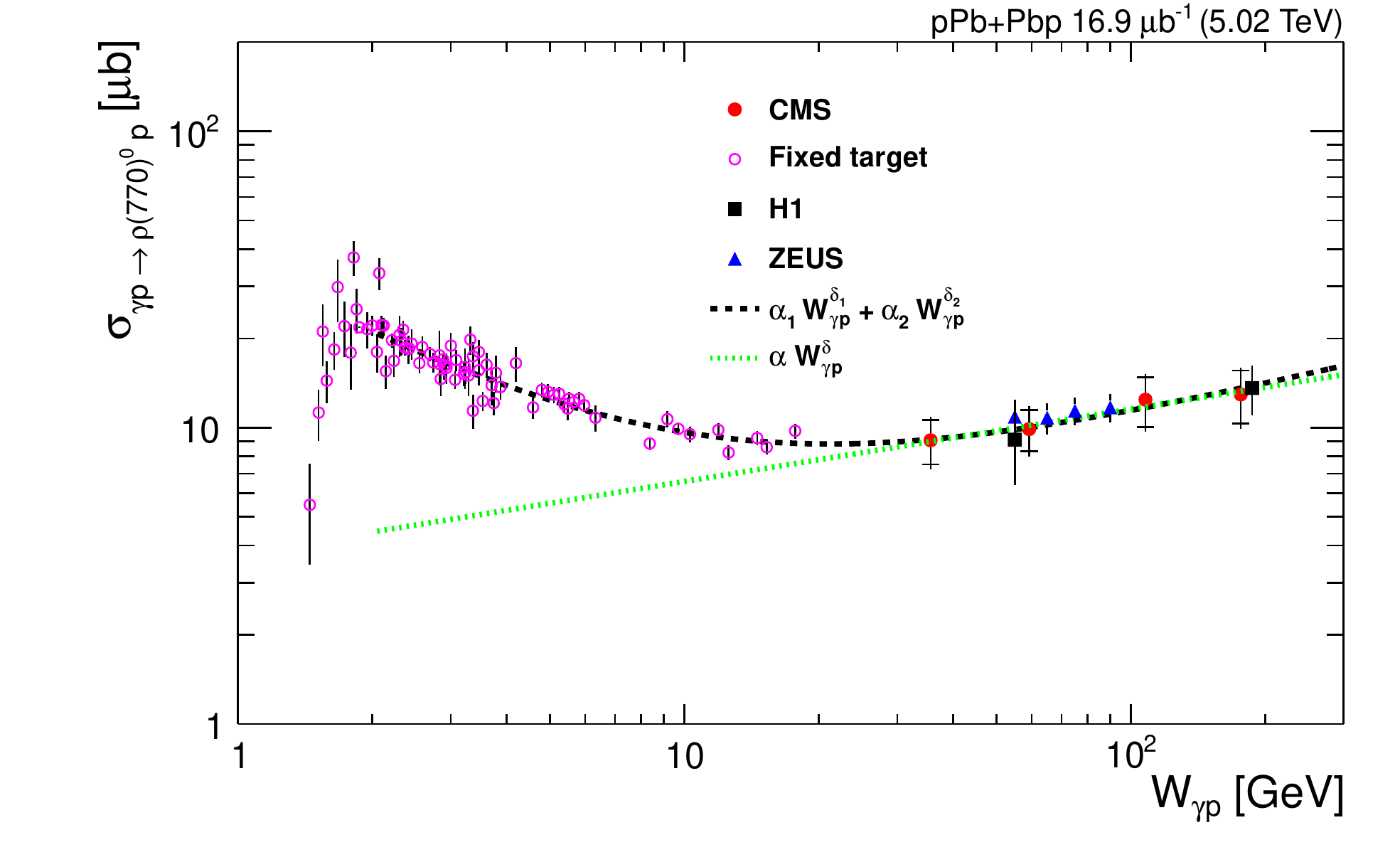}
\includegraphics[width=.45\textwidth]{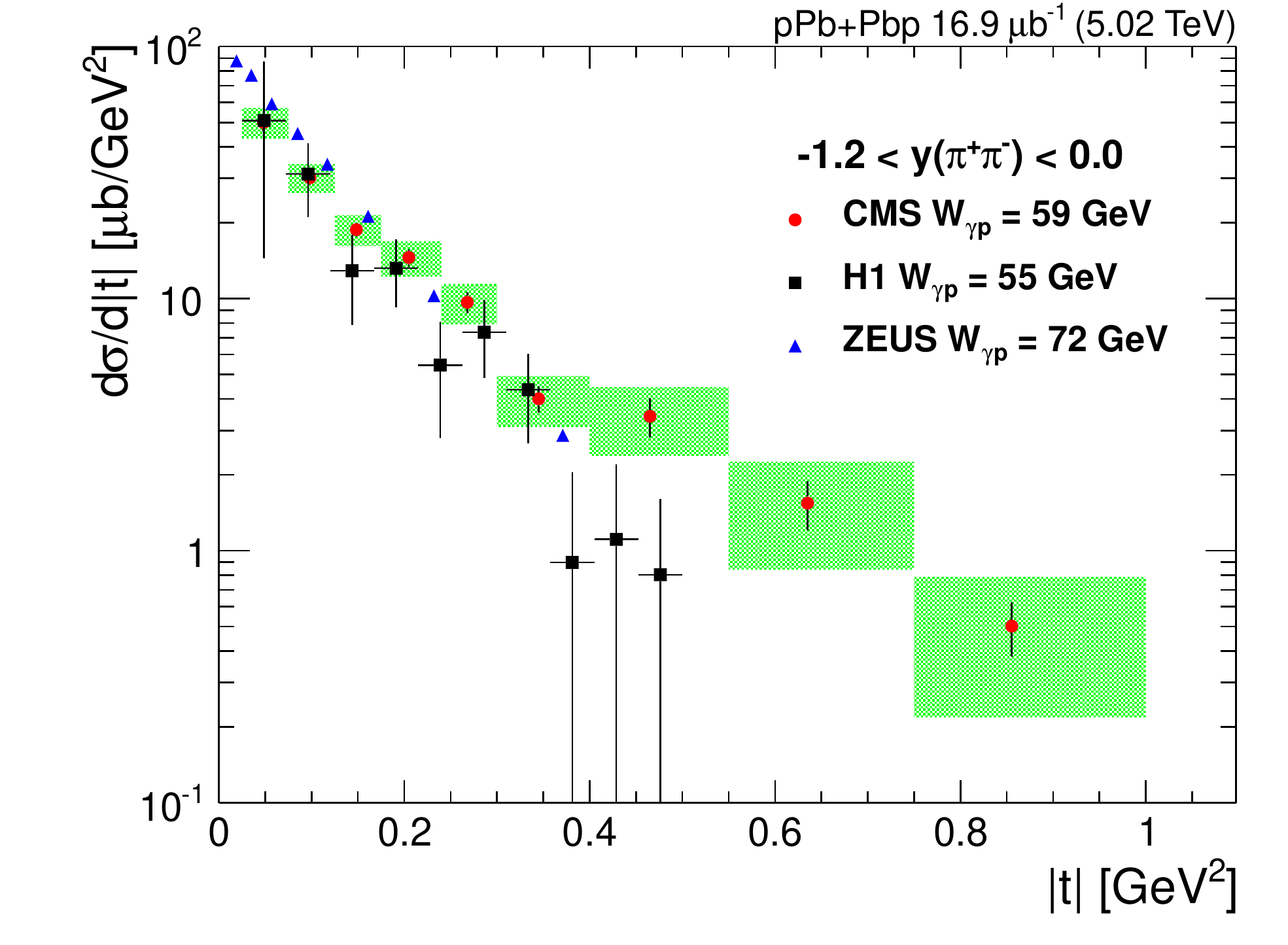}
\end{center}
\caption{\label{fig:rho} (Left) Photoproduction cross section as a function of the photon-proton center-of-mass energy, $W_{\gamma p}$. (Right) Number of events as a function of the transverse momentum of exclusive $\rho$ meson event candidates. Previous results by the H1 and ZEUS Collaborations are shown as well in both figures. Figures are extracted from Ref.~\cite{rho_CMS}.}
\end{figure}

Upon subtraction of background, just as with the aforementioned exclusive $\Upsilon$ production, one can estimate the four-momentum transfer at the proton vertex $-t \approx p_{T,\pi^+\pi^-}^2$, as shown in Fig.~\ref{fig:rho}. Additional studies related to the $b$ slope parameter extracted from the $d\sigma/d|t|$ in bins of $W_{\gamma p}$ are discussed in Ref.~\cite{rho_CMS}. Following a similar technique as with exclusive $\Upsilon$ production, one can extract the photoproduction cross section as a function of the photon-proton center-of-mass energy $W_{\gamma p}$. The main results of the latter are shown in Fig.~\ref{fig:rho}. The CMS results in both of these observables are consistent with similar studies done by the H1 and ZEUS experiments.

\section{Charged particle production in minimum bias events}\label{sec:mb}

Particle production without any selection bias arising from the requirement of the presence of a hard scattering process is known as ``minimum bias'' (MB). The bulk of these events occur at low momentum exchanges between the interacting partons inside the hadrons, where diffractive scattering or MPI play a significant role. Theoretical descriptions of these components of particle production are based on phenomenological models, whose parameters need to be ``tuned'' to reproduce experimental data. Given the large cross section of MB events in high energy pp collisions, it is of primary importance to have a good understanding of these processes, as they characterize properties of typical pileup interactions in each bunch crossing at the interaction point of CMS and other LHC experiments. One can characterize MB events by means of charged particle distributions. Charged particle distributions are measured for charged particles with $p_T > 0.5$ GeV and $|\eta|<2.4$ for events collected with a trigger selection MB events. The measured distributions are presented for different event data samples classifed by the calorimeter activity in the forward region. In this study, one considers the presence of at least one calorimeter tower with energy above $5$ GeV in $3<|\eta|<5$, and in some cases with a veto condition for towers with energy below $5$ GeV. The different event classes are as non-single diffractive enriched sample (NSD-enhanced) when there is calorimeter activity in both sides, as single diffractive enriched (SD-enhanced) when there is calorimeter activity on one side and a veto on the opposite side, and as inelastic when there is calorimeter activity on at least one side of CMS. The distribution labelled as SD-One-Side enhanced sample corresponds to the symmetrized distribution constructed from the SD-minus and SD-plus enhanced samples.

\begin{figure}[ht!]
\begin{center}
\includegraphics[width=.45\textwidth]{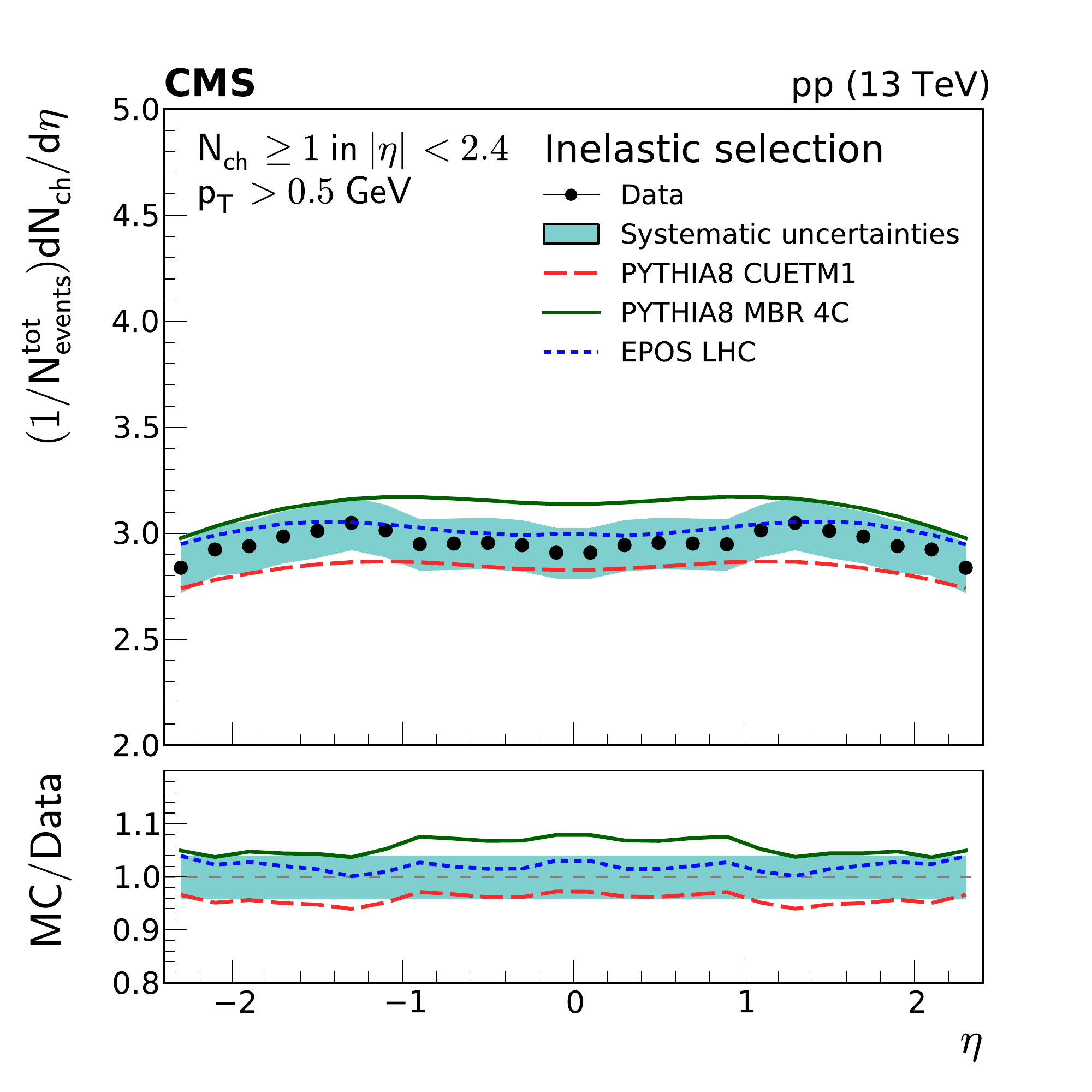}
\includegraphics[width=.45\textwidth]{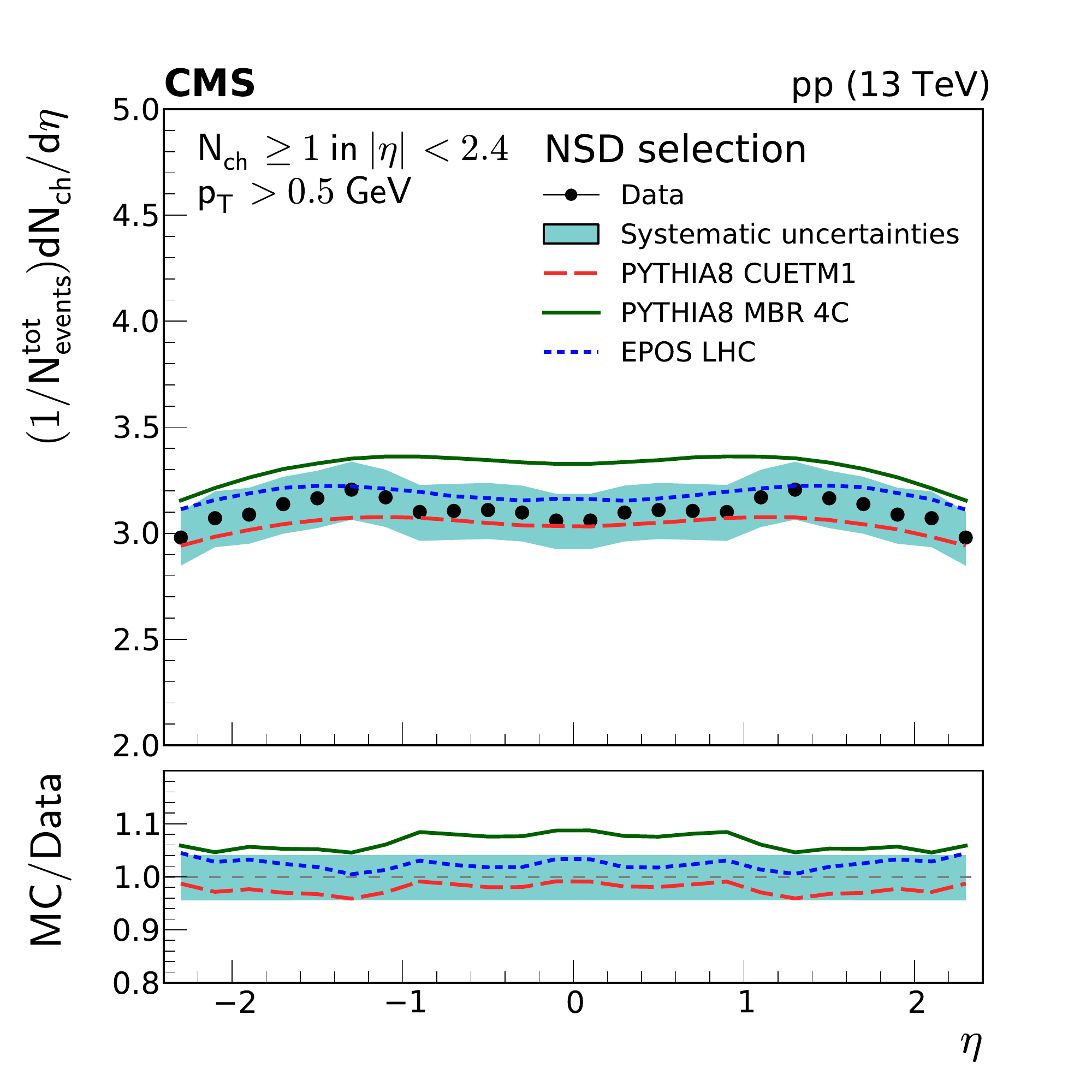}
\includegraphics[width=.45\textwidth]{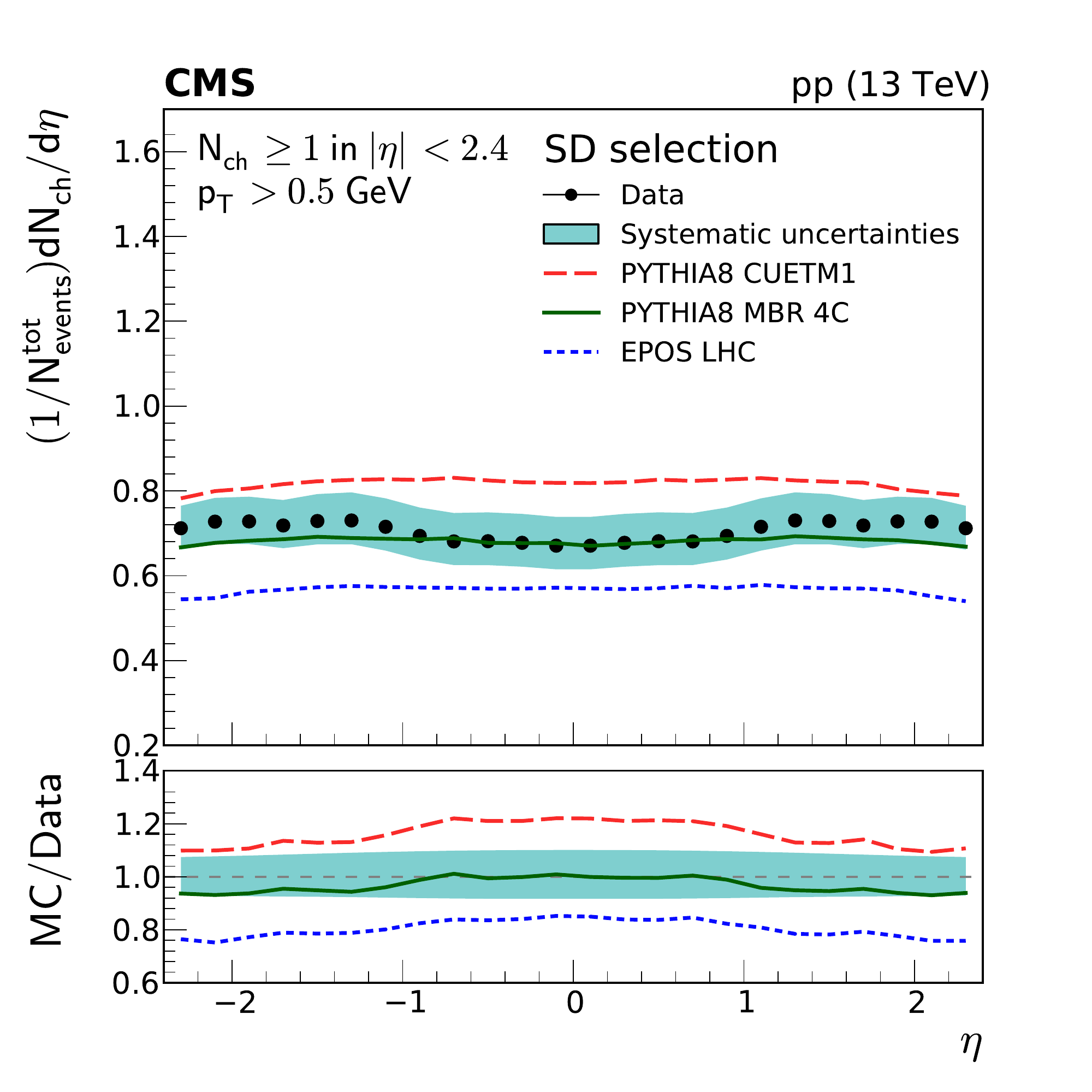}
\includegraphics[width=.45\textwidth]{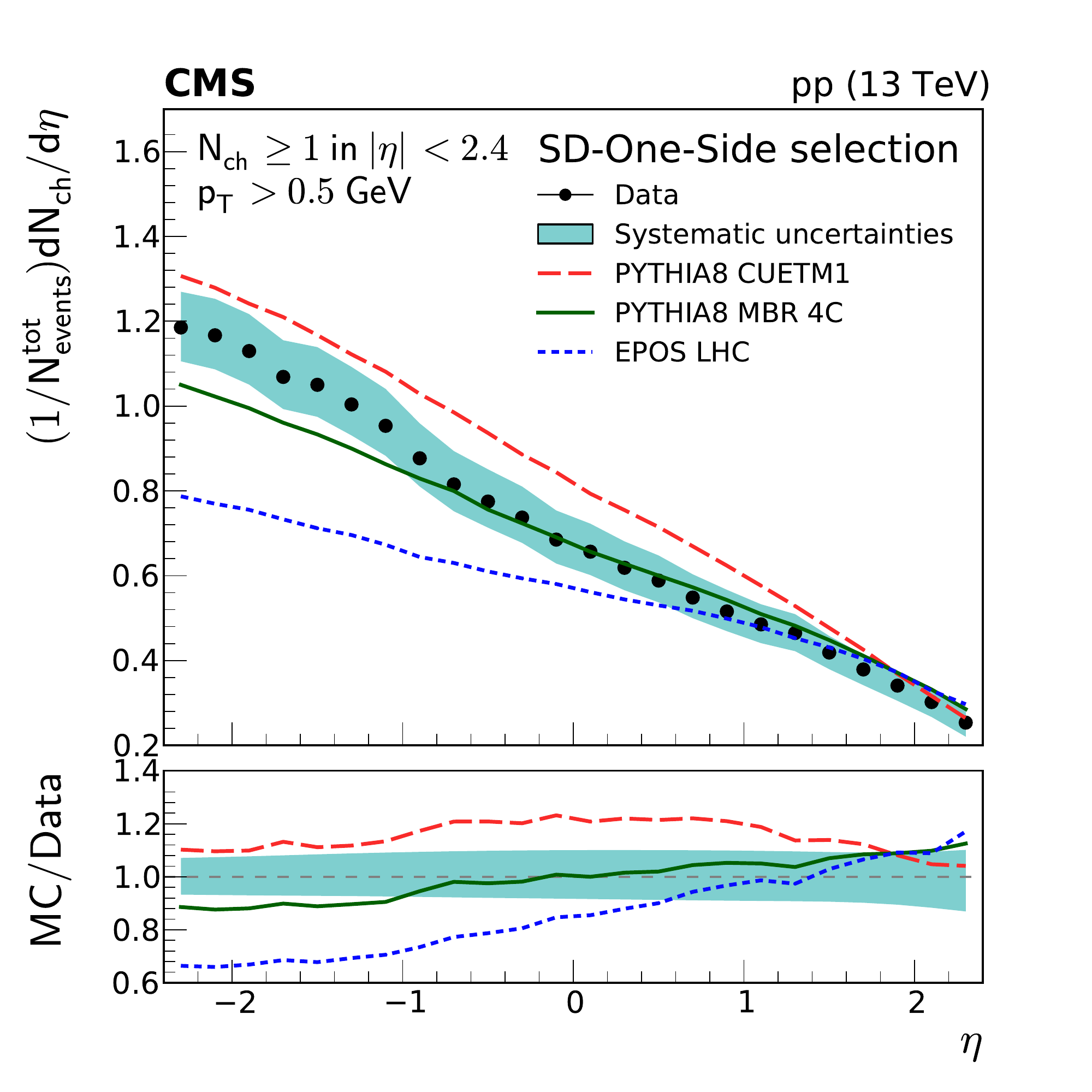}
\end{center}
\caption{\label{fig:minimumbias} Charged particle pseudorapidity densities averaged over both positive and negative $\eta$ ranges.  Top to bottom, left to right:  inelastic, NSD-, SD-, and SD-One-Side enhanced eventsamples. The measurements are compared to the predictions of the PYTHIA8 CUETM1 (long dashes), PYTHIA8 MBR4C (continuous line), and EPOS-LHC (short dashes) event generator. The error band represents the total systematic uncertainty.  Figures are extracted from Ref.~\cite{mb_CMS}.}
\end{figure}

The normalized particle distribution is measured as a function of the charged particle pseudorapidity for the four different selections, as shown in Fig.~\ref{fig:minimumbias}. The results are unfolded to particle level. PYTHIA8-CUETM1, PYTHIA8-MBR 4C~\cite{mbr}, and EPOS-LHC results are compared to the data. PYTHIA8 MBR 4C describes reasonably well the data for the SD-enhanced samples,but overestimates the yield in central pseudorapidities for the non-diffractive samples. PYTHIA8-CUETM1 and EPOS-LHC give a fair description for the non-diffractive samples, but they are off w.r.t.  data for the SD-enhanced selection. Additional observables based on the leading $p_T$ charged particle, and the per-event charged particle density, are shown in Ref.~\cite{mb_CMS}.

\section{Underlying event activity in Drell-Yan events at 13 TeV}

\begin{figure}[ht!]
\begin{center}
\includegraphics[width=.32\textwidth]{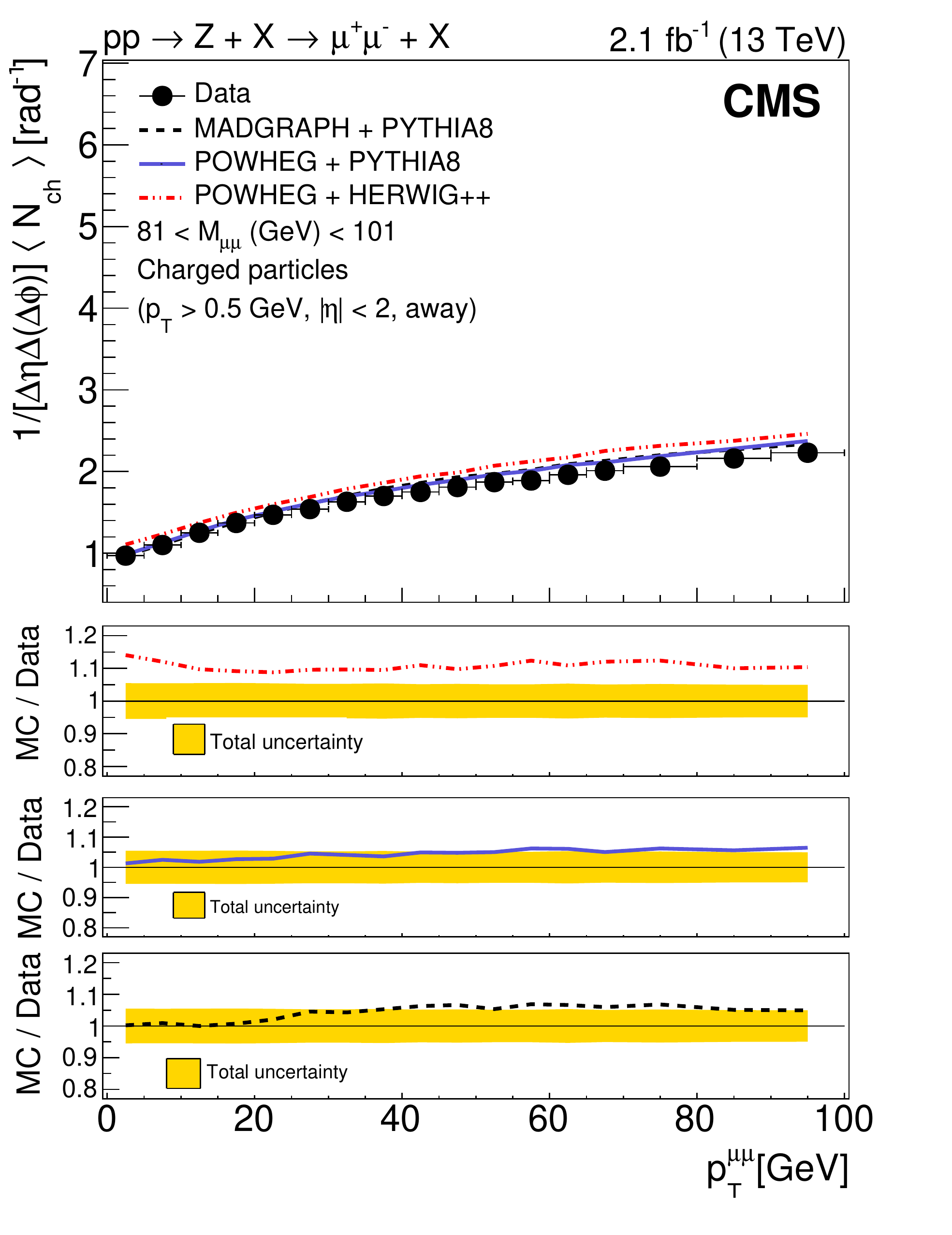}
\includegraphics[width=.32\textwidth]{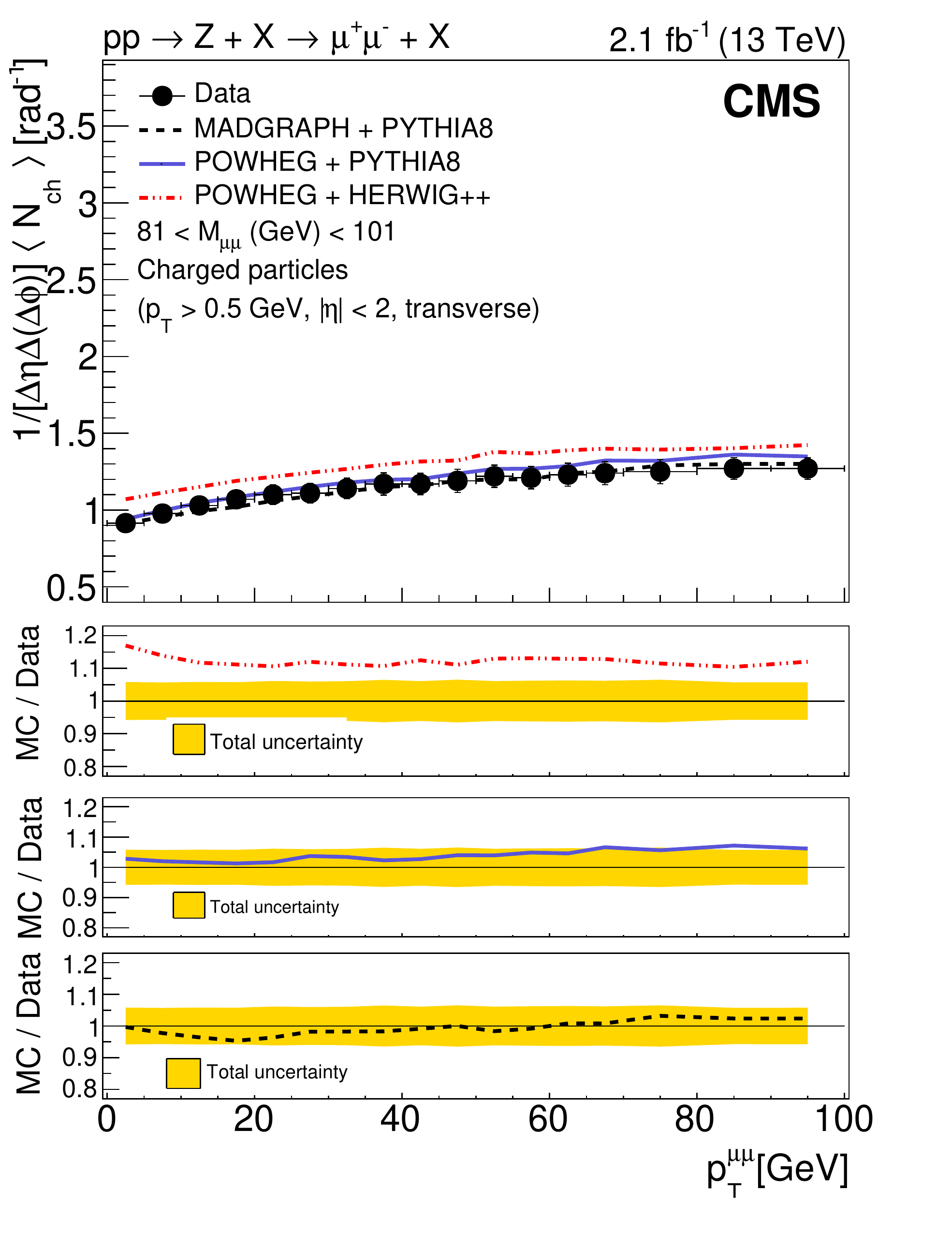}
\includegraphics[width=.32\textwidth]{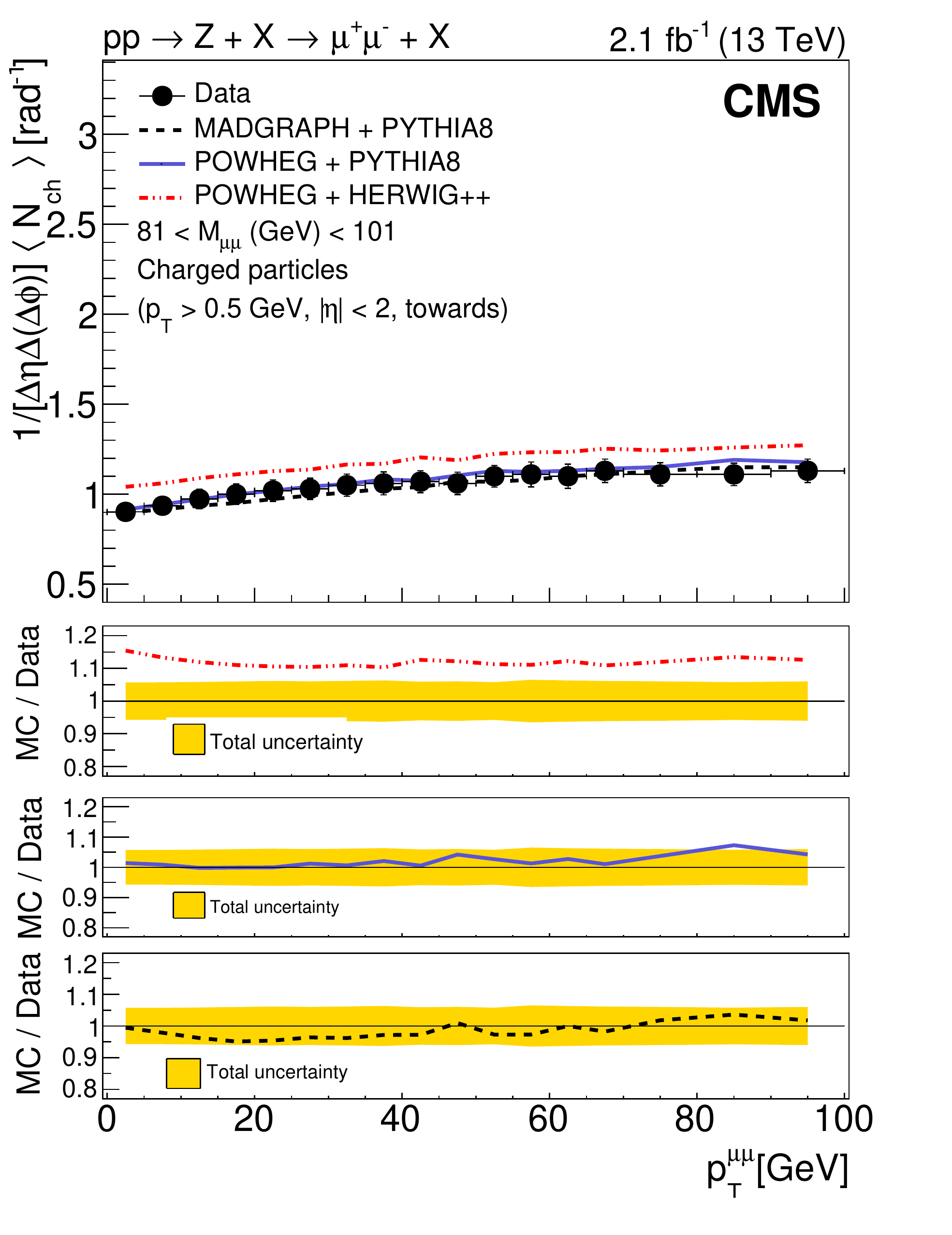}
\includegraphics[width=.32\textwidth]{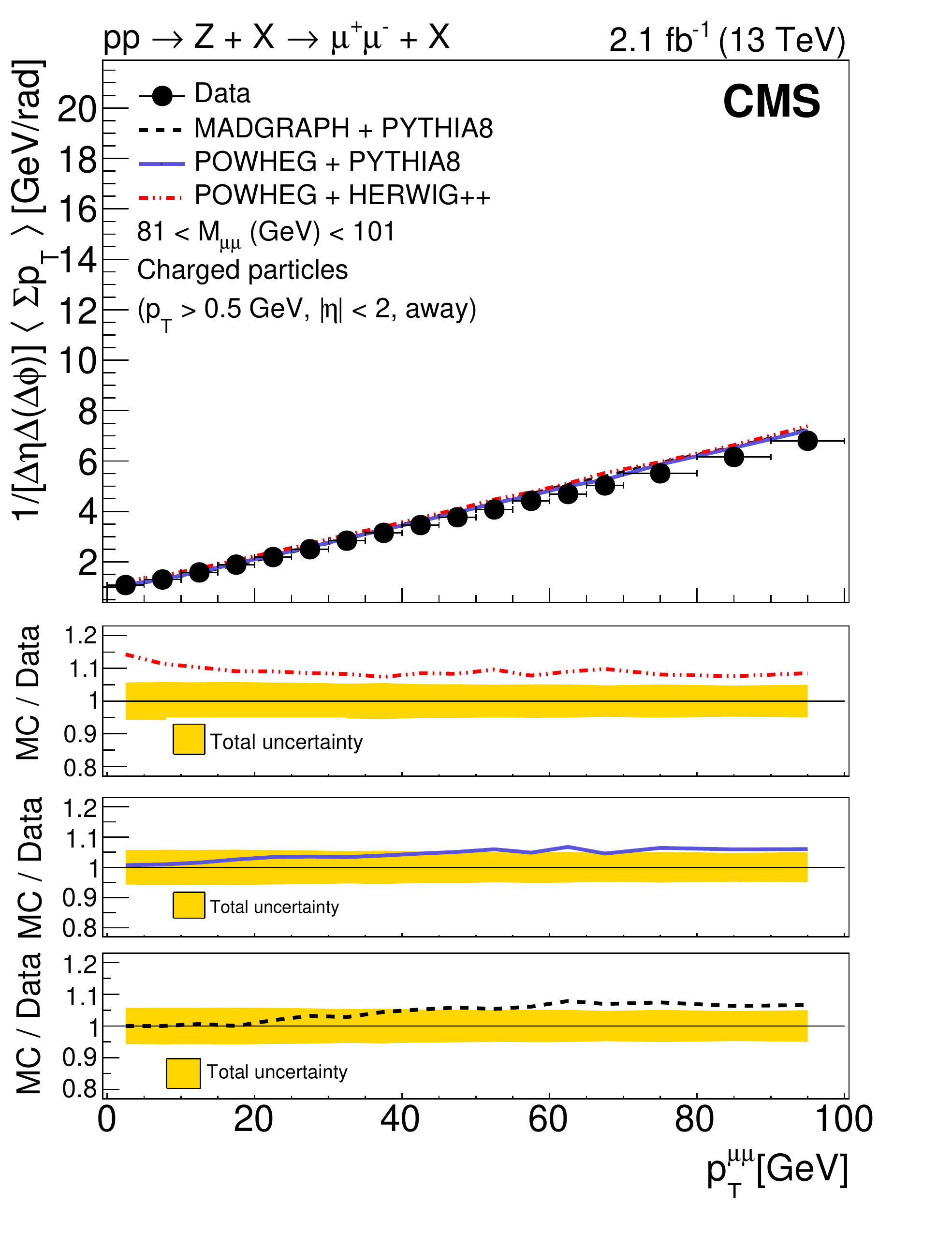}
\includegraphics[width=.32\textwidth]{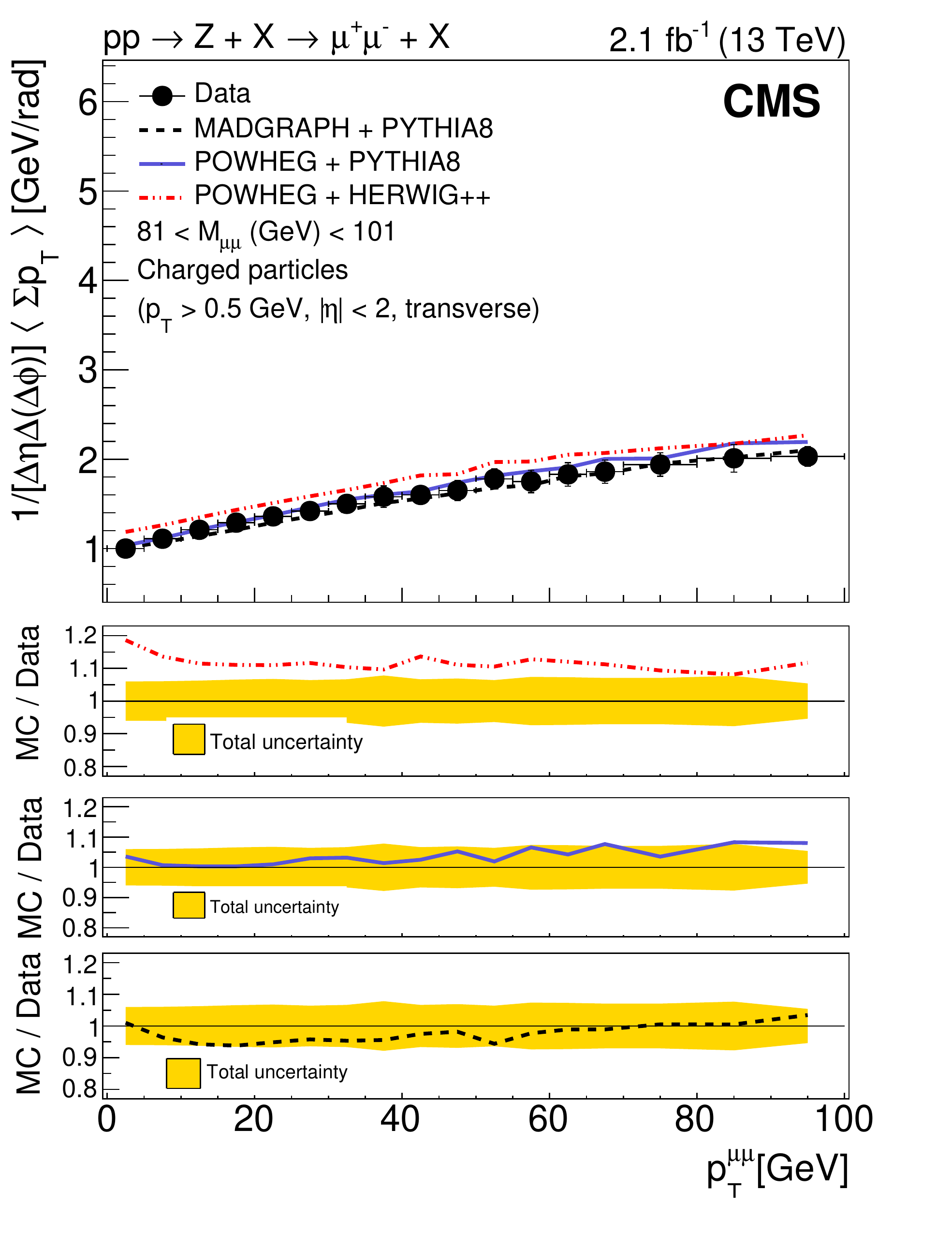}
\includegraphics[width=.32\textwidth]{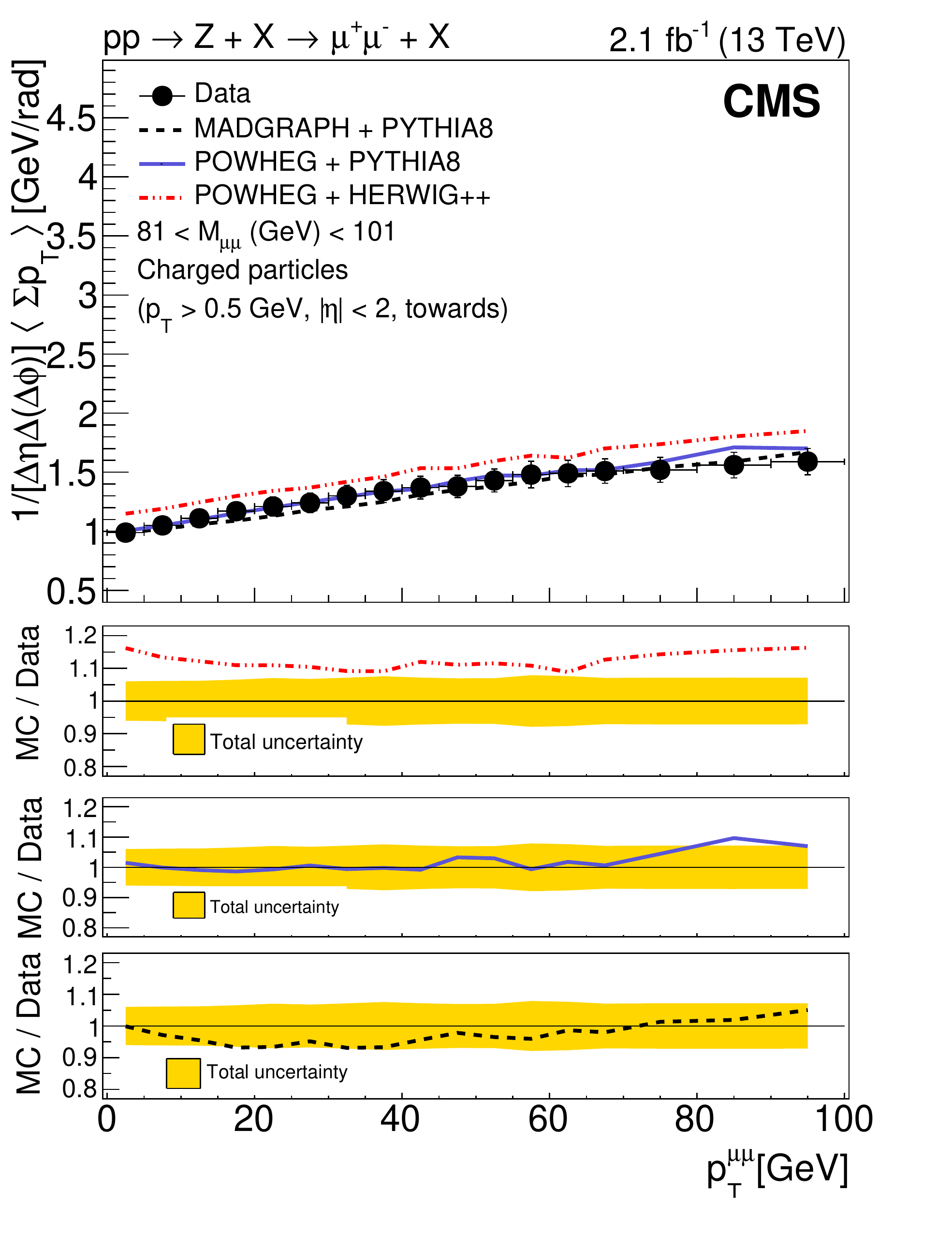}
\end{center}
\caption{\label{fig:average_energy_density} Particle  density (top) and $\Sigma p_T$ density (bottom) as a function of the dimuon transverse momenta in the away (left), transverse (center) and towards (right) regions. Figures are extracted from Ref.~\cite{ue_CMS}.}
\end{figure}

The underlying event (UE) activity is any activity stemming from beam-beam remnants and MPI. The UE produces particles carrying low transverse momentum, and are hard to disentangle from the initial-state radiation (ISR) and final-state radiation (FSR) present in the hard scattering process. The UE activity is usually quantified in terms of the charged particle multiplicity, as well as the scalar sum of the charged particles’ transverse momenta, in different angular regions defined with respect to a clean hard scattering process probe.

The CMS Collaboration presented a study of the UE activity based on 13 TeV pp collision data where the $Z$ boson ($pp\rightarrow Z+X$) is the hard scattering probe~\cite{ue_CMS}. In this study, the $Z$ boson decays into a $\mu^+\mu^-$ pair. This process is theoretically well understood, and has the additional advantage that FSR effets are not present. Muons are required to have $p_T^\mu>20$ GeV, $|\eta^\mu|<2.4$, and $81< m_{\mu\mu} < 101$ GeV.

The charged particle activity relative to the $Z$ boson direction is studied  in  three  angular  regions labelled as ``towards'', ``transverse'', and ``away'' regions, which are respectively  defined by $|\Delta\phi|<60^\circ$,  $60 < |\Delta\phi|<120^\circ$, and $|\Delta\phi|>120^\circ$, where $\Delta\phi$ is the azimuthal angle separation between the charged particle and the dimuon direction. Charged particles are requiredto satisfy $p_T>0.5$ GeV and $|\eta|<2$ for the measured distributions. The average particle density as a function of the dimuon transverse momentum is shown in Fig.~\ref{fig:average_energy_density}. Combinations of MADGRAPH and POWHEG (which include ISR effects for inclusive Z boson production) interfaced with PYTHIA8 and HERWIG++ for parton shower, hadronization effects, and UE activity, are compared to data. These combinations give reasonable agreement in shape w.r.t. the data, with the POWHEG+HERWIG++ off in normalization by $10$--$20$\%.

\section{Energy density as a function of pseudorapidity at 13 TeV}

A measurement of the energy density in MB events in proton-proton collisions $\sqrt{s}=13$ TeV within $-6.6 < \eta < -5.2$ and $3.15<|\eta|<5.20$ has been presented in Ref.~\cite{fwrd_energy_CMS}. Similar to the study of charged particle spectra in MB events, one is interested in characterizing soft particle production over a wide interval in pseudorapidity. In this study, the main focus is in forward pseudorapidities. The CASTOR calorimeter of CMS is used for negative pseudorapidities. The main observable in this study is the average energy density per collision, $dE/d|\eta|$. The energy density is extracted in three main event categories defined by the forward particle activity: inclusive inelastic (INEL) selection, non-single-diffractive-enhanced (NSD-enhanced), and single-diffractive-enhanced (SD-enhanced). Similar to the study described in Sec.~\ref{sec:mb}, these event categories are defined based on calorimeter energy deposits above noise level on at least one side (INEL), on both sides (NSD-enhanced), or in exactly one side (SD-enhanced). Comparisons are made to predictions based on PYTHIA8 + Monash2013, EPOS-LHC, and QGSJETII.

\begin{figure}[ht!]
\begin{center}
\includegraphics[width=.49\textwidth]{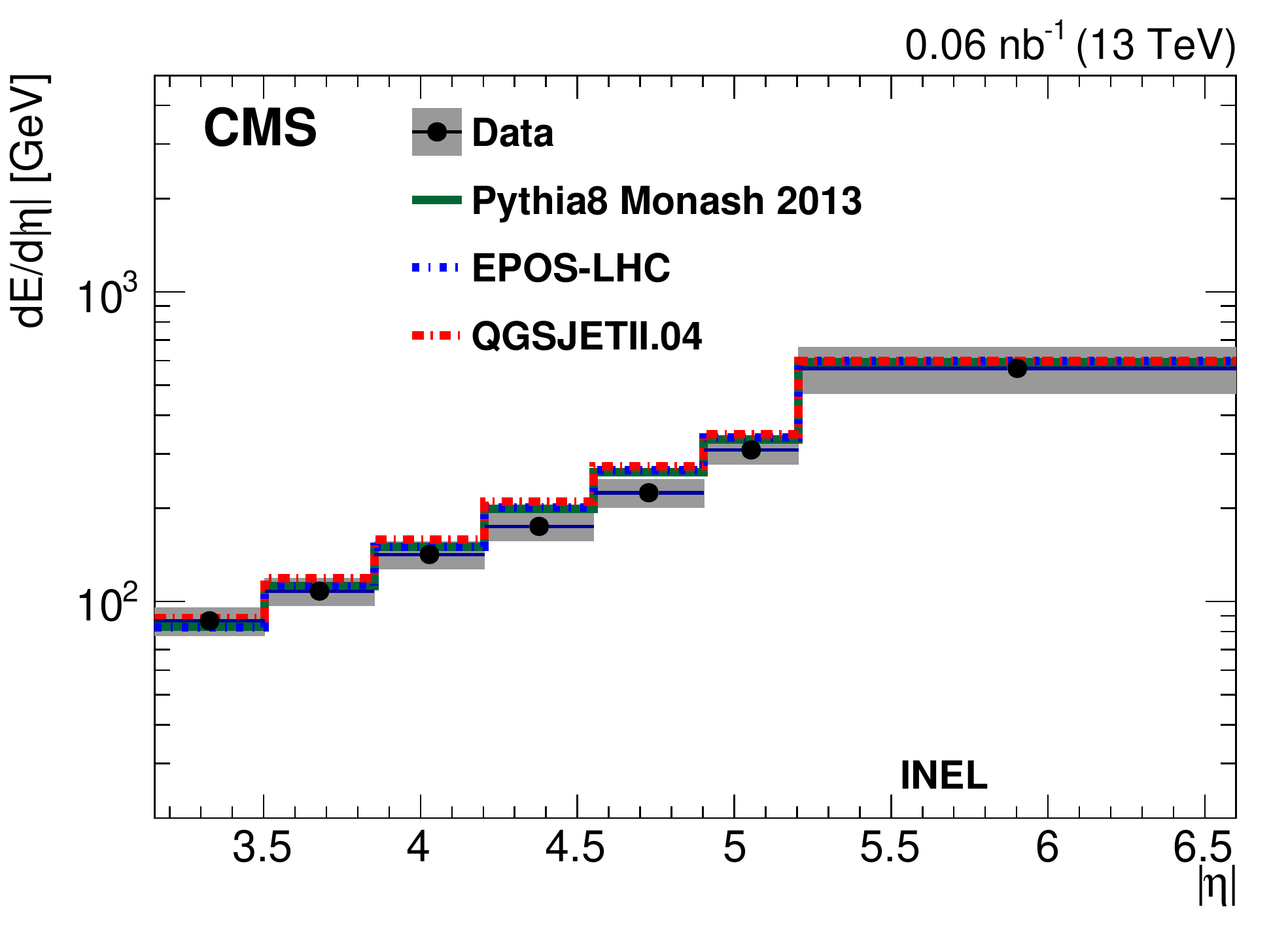}
\includegraphics[width=.49\textwidth]{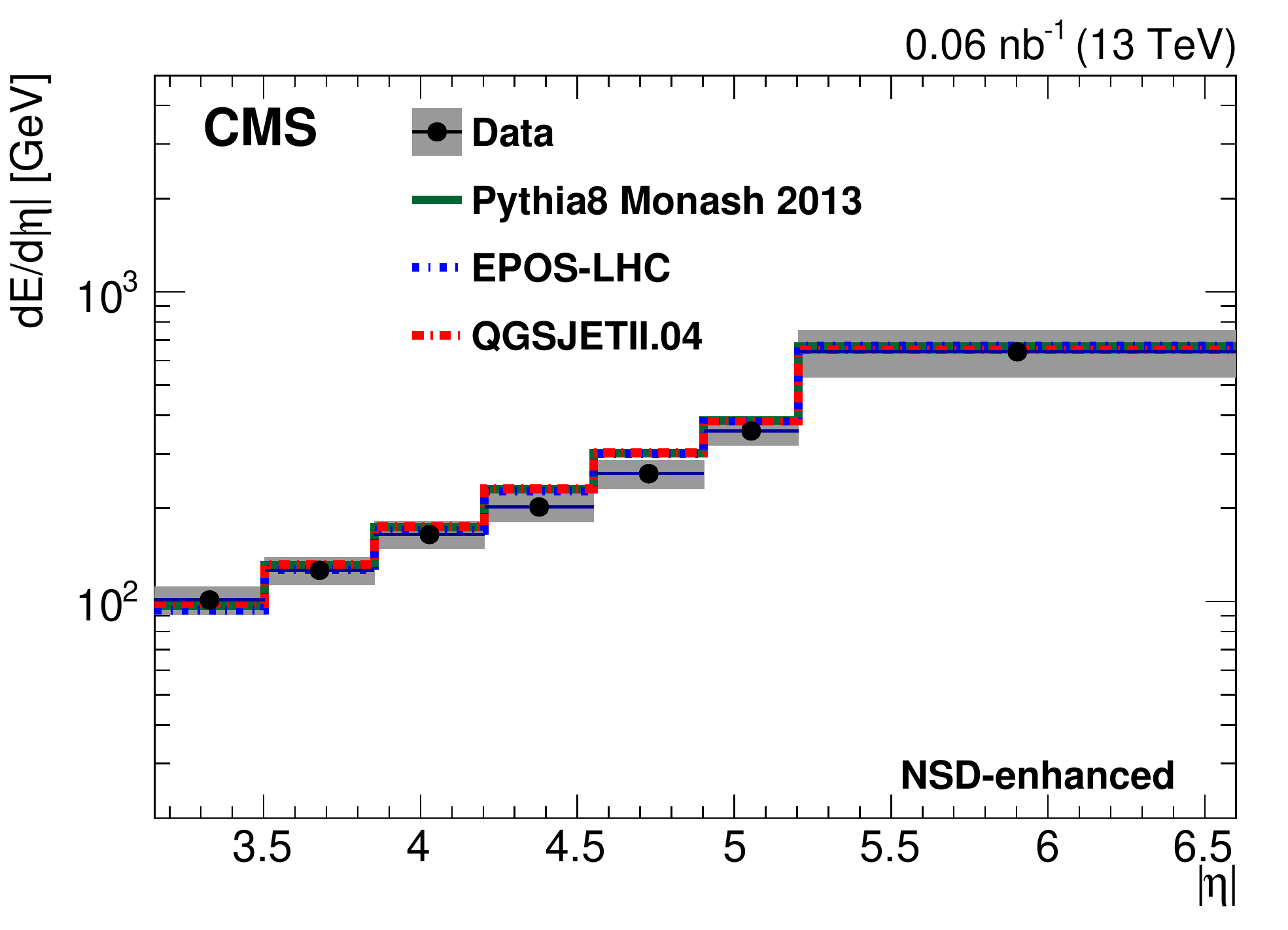}
\includegraphics[width=.49\textwidth]{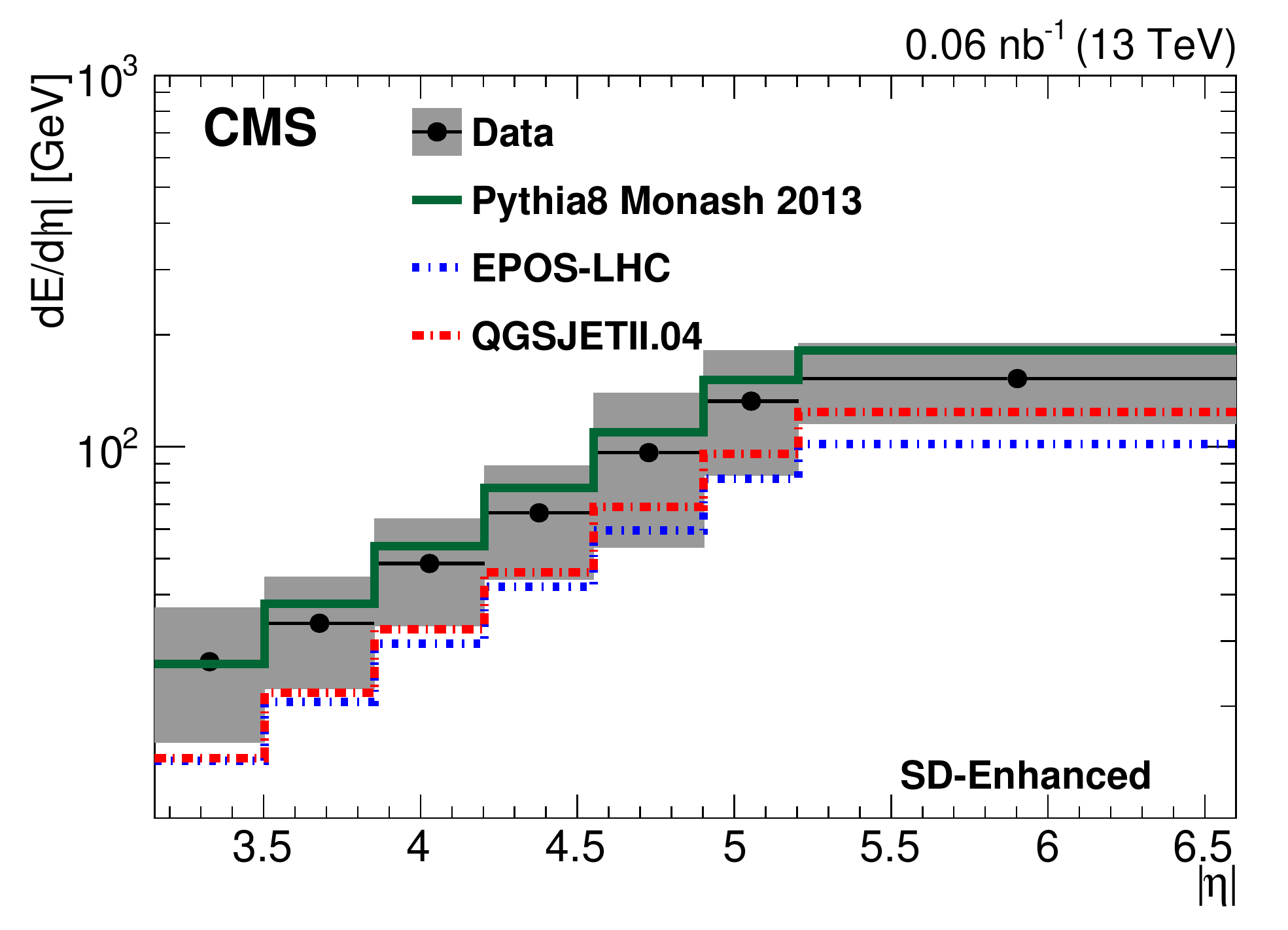}
\end{center}
\caption{\label{fig:forward_energy} Energy density at the stable-particle level for the INEL (Left), NSD-enhanced (Right), and SD-enhanced (Bottom) categories are compared to predictions from PYTHIA8 MONASH, EPOS-LHC, and QGSJETII.04. The gray band shows the total systematic uncertainty. Figures are extracted from Ref.~\cite{fwrd_energy_CMS}.}
\end{figure}

The resulting $dE/d|\eta|$ per category are shown in Fig.~\ref{fig:forward_energy}. The INEL and NSD-enhanced categories are extremely sensitive to MPI, while the SD-enhanced category is essentially unaffected. The Pythia8 Monash2013, EPOS-LHC, QGSJETII predictions overshoot the data for $|\eta| \approx 4.5$--$5$. The EPOS-LHC and QGSJETII models exhibit the largest differences when compared to the SD-enhanced results. A comparison with other PYTHIA8 tunes are described in Ref.~\cite{fwrd_energy_CMS}.

\begin{figure}[ht!]
\begin{center}
\includegraphics[width=.7\textwidth]{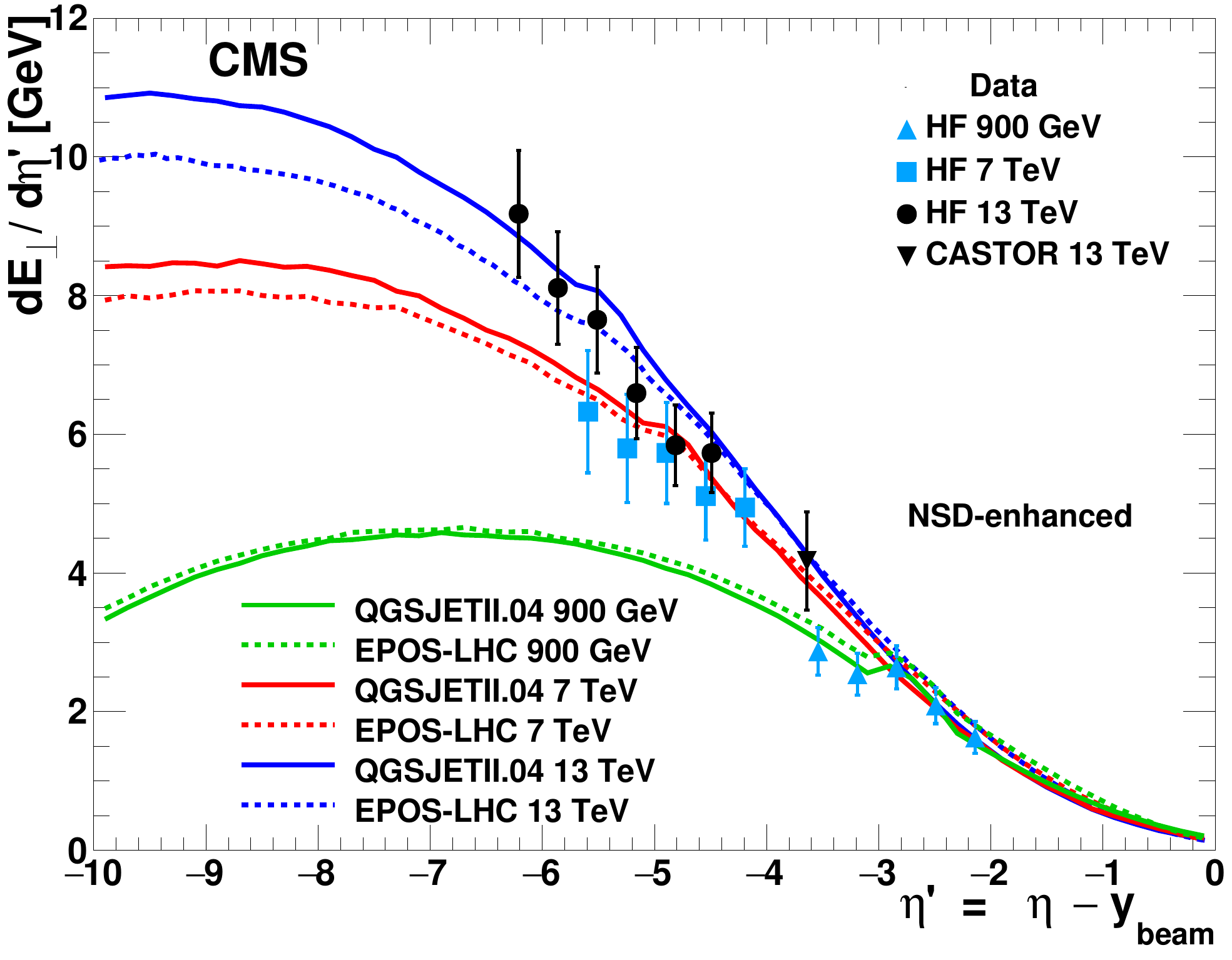}
\end{center}
\caption{\label{fig:limiting_fragmentation} A comparison of the measurements of transverse energy density  $\frac{d E_T}{\eta'}$  at 13 TeV, as a function of shifted pseudorapidity, $\eta'=\eta-y_\text{beam}$, to predictions for an NSD-enhanced selected sample at several different centre-of-mass energies. The error bars indicate the total systematic uncertainties. Figure is extracted from Ref.~\cite{fwrd_energy_CMS}.}
\end{figure}

At high energies the hypothesis of limiting fragmentation assumes a longitudinal scaling behaviour in terms of shifted pseudorapidity $\eta' = \eta - y_\text{beam}$, where $y_\text{beam}$ is the beam rapidity. Thus, soft-particle production in the projectile fragmentation region ($\eta' \approx 0$) is predicted to be independent of the center-of-mass energy. In this measurement, this is studied by measuring the transverse energy density $dE_T/d\eta$ with $E_T = E \cosh(\eta)$, and comparing it to measurements performed in pp collisions at different $\sqrt{s}$, as shown in Fig.~\ref{fig:limiting_fragmentation}. Predictions based on EPOS-LHC and QGSJETII models nicely describe the combined data in forward pseudorapidities, close to the projectile fragmentation region. Thus, the limiting fragmentation hypothesis is consistent with data. This is very important for the modelling of ultra-high energy interactions that typically occur in cosmic ray collisions.

\section{Summary}

Recent studies by the CMS experiment continue to shed light on regions of phase space highly relevant for the study of strong interactions at high energies. In these Proceedings, we have presented results by the CMS Collaboration related to the small-$x$ limit of QCD, where gluon densities are expected to grow rapidly due to multiple parton splitting within the proton or lead-nucleus, particle production in minimum bias events, and aspects related to multiparton interactions and beam-beam remnants.

Predictions based on BFKL calculations are consistent with results on azimuthal angle decorrelations in dijet events, where the two jets are separated by a large rapidity interval. However, predictions based on DGLAP evolution alone are also compatible with data for a wide range of rapidity. An extension of this study at higher energies and possibly with additional observables may help better disentangle BFKL dynamics from DGLAP dynamics. For events with two jets separated by a rapidity gap, we have the advantage that DGLAP dynamics is heavily suppressed, allowing for a description of jet-gap-jet events based on BFKL pomeron exchange. The challenge here is to simultaneously describe the short distance physics effects with the long distance physics effects (survival probability). A publication is in preparation regarding an extension of this study at $\sqrt{s} = 13$ TeV, which may help to better discriminate the existing models for jet-gap-jet event production.

Forward jet production in proton-nucleus collisions are very promising, and need to be further interpreted for stronger assessments on potential parton saturation effects observed in data. The results in exclusive quarkonia production in proton-lead collisions demonstrate the feasibility of continuing the program first started by the HERA experiments. An extension of these studies at larger energies and larger samples in proton-lead collision data collected in 2016 would help constrain better the gluon density in the small-$x$ region.

Measurement of MB events and UE activity are sensitive to the various mechanisms for particle production in proton-proton collisions with increasing precision at various center-of-mass energies at the CERN LHC. The latter provide very important inputs and constraints for MC generator tuning necessary for collider searches and cosmic-ray physics studies.

\section*{Acknowledgements}

I would like to thank the U.S. Department of Energy for their generous support (grant number DE-SC0019389.). I would also like to thank Universidad de Guanajuato for hosting and organizing this event.

\bibliographystyle{JHEP} 
\bibliography{references.bib}

\printindex

%IEEM
%MBR
%EPOS-LHC

\end{document}